\documentclass[fleqn,usenatbib,twocolumn, linenumbers]{mnras}

\usepackage{newtxtext,newtxmath}

\usepackage[T1]{fontenc}

\DeclareRobustCommand{\VAN}[3]{#2}
\let\VANthebibliography\thebibliography
\def\thebibliography{\DeclareRobustCommand{\VAN}[3]{##3}\VANthebibliography}

\usepackage{graphicx}	
\usepackage{amsmath}	
\usepackage{subfigure}
\usepackage{verbatim}
\usepackage{color}
\usepackage[figuresright]{rotating}
\usepackage{longtable}
\usepackage{threeparttable}
\usepackage{multirow}

\title[SN 2022acko]{SN 2022acko: a low-luminosity SNe IIP with signs of early circumstellar interaction}

\author[Lin et al.]{
Han Lin,$^{1,2,3}$
Jujia Zhang,$^{1,3,4}$\thanks{E-mail:jujia@ynao.ac.cn}
Xiaofeng Wang,$^{5}$
Maokai Hu,$^{5}$
Shuai Zha,$^{1,3}$
Danfeng Xiang,$^{5,6}$
Liping Li,$^{1,3}$
\newauthor
Andrea Reguitti,$^{7,8}$
Xinghan Zhang,$^{9}$
Yongzhi Cai,$^{1,3,4}$
Zhenyu Wang,$^{1,3,10}$
Zeyi Zhao,$^{1,3,10}$
Qian Zhai,$^{1,4}$
\newauthor
Fang Huang,$^{11}$
Weili Lin,$^{12}$
Jinming Bai,$^{1,3,4}$
\\
$^{1}$Yunnan Observatories, Chinese Academy of Sciences, Kunming 650011, China\\
$^{2}$Key Laboratory of Radio Astronomy and Technology, Chinese Academy of Sciences, A20 Datun Road, Chaoyang District, Beijing, 100101, P. R. China\\
$^{3}$International Centre of Supernovae, Yunnan Key Laboratory, Kunming 650216, P. R. China\\
$^{4}$Key Laboratory for the Structure and Evolution of Celestial Objects, Chinese Academy of Sciences, Kunming 650011, China\\
$^{5}$Physics Department, Tsinghua University, Beijing, 100084, China\\
$^{6}$Beijing Planetarium, Beijing Academy of Science and Technology, Beijing 100044, China \\
$^{7}$INAF - Osservatorio Astronomico di Brera, Via E. Bianchi 46, 23807 Merate (LC), Italy \\
$^{8}$INAF - Osservatorio Astronomico di Padova, Vicolo dell'Osservatorio 5, 35122 Padova, Italy \\
$^{9}$School of Physics and Electronic Information, Jiangsu Second Normal University, Nanjing, Jiangsu 211200, PR China\\
$^{10}$School of Astronomy and Space Science, University of Chinese Academy of Sciences, Beijing, 101408, China\\
$^{11}$Department of Astronomy, Shanghai Jiao Tong University, Shanghai 200240, China\\
$^{12}$Department of Astronomy, Xiamen University, Xiamen, Fujian 361005, China\\
}

\date{Accepted XXX. Received YYY; in original form ZZZ}

\pubyear{\the\year{}}

\begin{document}
\label{firstpage}
\pagerange{\pageref{firstpage}--\pageref{lastpage}}
\maketitle

\begin{abstract}
We present optical-ultraviolet photometry and optical spectra for the type II supernova (SN) 2022acko. 
The spectroscopic observations span phases from $\sim$ 1.5 to $\sim$ 60 days after the explosion, while the light curve was observed up to $\sim$ 300 days. 
The V-band peak is $-15.5 \pm 0.3$ mag, suggesting that SN 2022acko is a low-luminosity SN II (LLSN). 
The 
overall observed properties of SN 2022acko are consistent with those produced by a lower mass progenitor ($\rm M_{ZAMS} \sim $9-10M$_{\odot}$).
The spectra at $t=1.5$\,d and $t=2.5$\,d exhibit a broad emission feature peaking near 4600 \AA \ (the ``ledge'' feature), which we interpret as blueshifted He~{\sc ii} 4686 \AA \ lines arising from the ionized ejecta. 
Moreover, a possible flash-ionized (FI) emission line of H$\alpha$ (FWHM $\sim 1100\ \rm km \ s^{-1}$) was superposed on the broad emission component of H$\alpha$ P-Cgyni profile in the $t=1.5$\,d spectrum. 
Assuming an ejecta velocity of $\rm 12000\ km\ s^{-1}$, the rapid disappearance of this narrow H$\alpha$ emission line within two days suggests highly confined CSM within $\sim \rm 2\times10^{14}\, cm$. 
Assuming a spherically symmetric CSM, the mass loss rate within this radius is estimated to be $\rm \sim 5 \times 10^{-4} M_{\odot} \ year^{-1}$ based on our hybrid light curve model.
The early ``ledge'' feature observed in SN 2022acko have also been observed in other SNe II, suggesting that early-phase circumstellar interaction (CSI) is more common than previously thought. 
\end{abstract}





\section{Introduction}

Type II supernovae (SNe II), characterized by prominent hydrogen lines in their spectra, are produced by the core collapse of massive stars (>8 $\rm M_{\odot}$) \citep{1997ARA&A..35..309F,2009ARA&A..47...63S}. 
The most common subtypes of SNe II are SNe IIP and IIL.
Although SNe IIP and SNe IIL exhibit different light curve shapes, their photometric and spectral properties show a continuous distribution with SN IIL showing faster declining light curves, brighter luminosity, higher expansion velocities and shallow H$\alpha$ absorption \citep{2014ApJ...786...67A,2014ApJ...786L..15G,2015ApJ...799..208S,2016AJ....151...33G,2016MNRAS.459.3939V,2019MNRAS.490.2799D,2024MNRAS.528.3092L}. 
Low-luminosity SNe (LLSNe), which are located at the low luminosity end of whole IIP/L sample, generally exhibit lower expansion velocities, smaller explosion energies, and reduced amounts of nickel synthesized in the explosion \citep{2004MNRAS.347...74P,2014MNRAS.439.2873S}. The prototype of this class is SN 1997D \citep{1998ApJ...498L.129T,2001astro.ph.11573B}. Two different scenarios have been proposed for its origin, i.e., the core-collapse explosion of a low-mass red supergiant (RSG) \citep{2000A&A...354..557C}, or a low-energy explosion of a high-mass star with a large amount of fallback \citep{1998ApJ...498L.129T}.

In recent years, both hydrodynamical modelling of observables \citep{2014MNRAS.439.2873S,2022MNRAS.514.4173K}, and constraints from pre-SN archival images, such as those of SN 2005cs \citep{2005MNRAS.364L..33M,2006ApJ...641.1060L} and SN 2008bk \citep{2012AJ....143...19V}, 
suggest that LLSNe originate from low to intermediate-mass RSGs.
In addition, the progenitor mass range of LLSNe overlaps with that of the electron-capture (EC) explosions theoretically predicted for super-asymptotic giant branch (AGB) stars \citep{1984ApJ...277..791N}. This potential overlap makes LLSNe important candidates for theoretical electron-capture supernovae (ECSNe) \citep{2018ApJ...861...63H,2022MNRAS.513.4983V}.

Massive stars lose mass throughout their evolution, and when an SN explodes in the presence of circumstellar material (CSM), the interaction between the SN ejecta and the CSM can be used to diagnose the circumstellar environment and the mass-loss history of the progenitor. With the rapid, high-cadence optical transient surveys, an increasing number of SNe II were discovered within a few days, even hours after their explosions. 
A significant fraction ($>$ 30\%) \citep{2021ApJ...912...46B,2023ApJ...952..119B} of SNe II exhibit Flash-Ionized (FI) features in their early spectra, characterized by narrow emission lines with broad electron-scattering wings. 
The FI features are produced by the recombination of the unshocked CSM which was ionized by high-energy photons generated during shock breakout or CSM interaction (CSI) \citep{2000ApJ...536..239L, 2014Natur.509..471G,2015MNRAS.449.1876S,2016ApJ...818....3K}.
Once the CSM was swept up by the SN ejecta, the FI features disappear. It is suggested that progenitors exhibiting FI features possess a dense, confined CSM environment, indicative of violent mass loss shortly before the SN explosion \citep{2016ApJ...818....3K,2021ApJ...912...46B}. 

Different from the standard FI features, the early spectra of a few SNe II show a broad emission peaked near 4600\AA , known as the ``ledge'' feature \citep{2020ApJ...902....6S}, which may also indicate weak or ambiguous signatures of CSM interaction.
The ledge feature has been observed in several SNe II, including SN 2006bp \citep{2007ApJ...666.1093Q}, SN 2017eaw \citep{2019MNRAS.485.1990R}, SN 2021yja \citep{2022ApJ...935...31H}. 
Though interpretations of the ``ledge'' feature vary, a common interpretation is that it results from an RSG Progenitor with an extended envelope exploding into low-density CSM \citep{2022ApJ...935...31H,2023ApJ...945..107P}.

In this work, we present observations and analysis of a LLSN 2022acko \citep{2022TNSCR3549....1L} that exhibits the ``ledge'' feature due to early circumstellar interaction (CSI), contributing to the rare class of LLSNe (i.e., SN 2016bkv \citep{2018ApJ...859...78N,2018ApJ...861...63H}, SN 2018lab \citep{2023ApJ...945..107P} and SN 2021gmj \citep{2024MNRAS.528.4209M,2024ApJ...971..141M}) with observational evidence of short-lived CSM interaction. 
As a nearby LLSN with possible CSI, a series of unique observations were obtained for SN 2022acko.
For the first time, the James Webb Space Telescope (JWST) was used to identify its candidate progenitor \citep{2023MNRAS.524.2186V}.  \citet{2023MNRAS.524.2186V} suggested that the candidate progenitor star has an initial mass somewhat less than the 8\,$\rm M_{\odot}$, however, the SED and luminosity of the candidate are inconsistent with that of a super-asymptotic giant branch star that might produce an electron-capture SN. 
\citet{2024arXiv240114474S} presented JWST spectroscopic and photometric observations of SN 2022acko at $\sim$ 50 days past explosions, marking the first JWST spectroscopic observations of a core-collapse SNe (CCSNe).  
The observation of SN 2022acko set a baseline for future JWST observations that aim to see molecular and dust formation. 
\citet{2023ApJ...953L..18B} presented the first early Far-ultraviolet spectra for this object, identified the ions that
dominate the early ultraviolet spectra, and created an ultraviolet spectral time-series of SNe II.

In Section 2, we describe the observations, data reduction, and the basic information of SN 2022acko. The photometric and spectroscopic evolution of SN 2022acko are presented in Sections 3 and 4, respectively. We discuss the signatures of CSM interaction, and explosion parameters of SN 2022acko in Section 5 and summarize our conclusions in Section 6.

\section{OBSERVATIONS AND DATA REDUCTION}

\subsection{Discovery}
\begin{figure}
	\includegraphics[width=0.9\columnwidth]{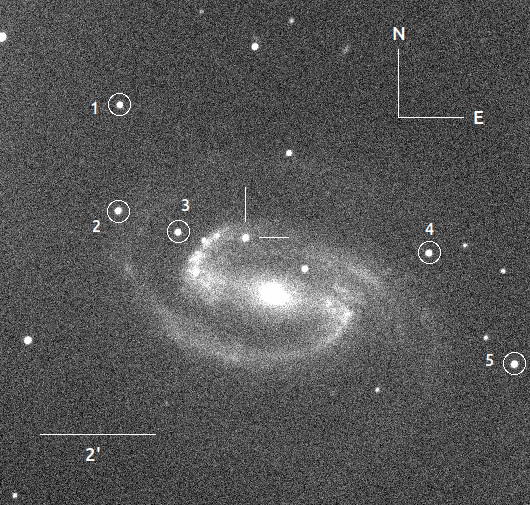}
	\caption{$g$-band image of SN 2022acko in NGC 1300, taken on 2022 Dec. 15, with the Lijiang 2.4-m telescope. North is up and east is to the right. The locations of the supernova and local reference stars are marked by white tick marks and circles, respectively.}
	\label{fig:2022acko}
\end{figure}

SN 2022acko was discovered on UT 2022-12-06 02:11:30 (JD 2459919.5913) by the DLT40 team \citep{2018ApJ...853...62T}, with a clear band magnitude of 16.54 mag \citep{2022TNSTR3543....1L}. Its J2000 coordinates are RA = 03:19:38.990 and DEC = -19:23:42.68, located at $29^{\prime \prime}.8\,$west and $58^{\prime \prime}.0\,$north of the center of host galaxy NGC 1300 (See Figure \ref{fig:2022acko}). A non-detection on JD 2459918.6726 with a 5$\sigma$ upper limit of 18.51 mag (o-band) was provided by ATLAS forced photometry. We adopt this non-detection point as the estimated explosion time of SN 2022acko to facilitate comparison with \citet{2023ApJ...953L..18B} and \citet{2023MNRAS.524.2186V}.
SN 2022acko was classified as a young SN by \citet{2022TNSCR3549....1L} based on the spectrum obtained from the Lijiang 2.4m telescope (LJT; \citealp{2015RAA....15..918F}),  which displays broad emission features at 4600 \AA. Extensive follow-up observations have been carried out for this object.

The Milky Way (MW) reddening for SN 2022acko is $E(B-V)_{\rm MW} = 0.026$ \citep{2011ApJ...737..103S}, corresponding to a V-band extinction of 0.083 mag assuming $R_{V}$ =3.1. 
\citet{2010ApJ...715..833O} and \citet{2009ApJ...694.1067P} suggested that the SN host galaxy may follow a Low$-R_{V}$ extinction law compared to the Milky Way. Adopting $R_V = 1.4$ for host galaxy and $E(B-V)_{\rm host} = 0.03 \pm 0.01$\,mag (measured by \cite{2023ApJ...953L..18B} from Na~{\sc i} D lines), the host extinction for V-band is found to be $A_{\rm V, host}$=0.04 $\pm$ 0.01\,mag and the total extinction for V-band is therefore $A_{\rm V, total}$ = 0.12 $\pm$ 0.01\,mag. 
The basic information of SN 2022acko is listed in Table \ref{tab:B_I}. 

\subsection{Distance}
The average Tully-Fisher distance (from the NASA Extragalactic Database (NED)\footnote{https://ned.ipac.caltech.edu/}) for the host galaxy of SN 2022acko is 14.99 $\pm$ 4.30 Mpc. 
In previous work on SN 2022acko (i.e.,\citealt{2023ApJ...953L..18B, 2023MNRAS.524.2186V}), 
a distance of $18.99\pm2.9$ Mpc, which derived from the PHANGS
survey \citep{2021MNRAS.501.3621A} using the Numerical Action Method \citep{2017ApJ...850..207S}, was adopted. 
We noticed that there is a large difference between these two methods, and the distance will largely affect the luminosity, which will affect the classification and other derived properties. 
We therefore derive the distance to SN 2022acko using the expanding photosphere method (EPM; \citealt{1974ApJ...193...27K} and the standard candle method (SCM; \citealt{2002ApJ...566L..63H}), adopting the average value as the final distance.

EPM assumes that the SN radiates as a diluted blackbody in early phases, which:
\begin{equation}
    4\pi \xi ^2 R_{\rm ph}^2 \pi B_{v}(T) = 4\pi D^{2} f_v^{\rm dered}
\end{equation}
where $\xi$ is the dilution factor, $ R_{\rm ph}$ is the photospheric radius, $B_{v}(T)$ is the Planck function at temperature $T$, $D$ is the distace to SN, and $f_v^{\rm dered}$ is the de-reddened observed flux.
The distance is derived by relating the apparent
angular radius $\theta$ of the photosphere to its physical radius $R_{\rm ph} = v_{\rm phot}(t-t_0)$:
\begin{equation}
    D = R_{\rm ph}/\theta = v_{\rm phot}(t-t_0)/\theta
\end{equation}
where $v_{\rm phot}$ is the photospheric velocity at time $t$, and $t_0$ is the explosion epoch.
It is a geometrical technique that can independently constrain the distance of SNe II.
In this work, we first determined the temperature and $\xi^2 \theta^2$ by fitting the blackbody function to the multiband photometry. 
The combination of UV and optical bands was used at epochs earlier than 3 days after the explosion, while only the optical bands were used later than 3 days after the explosion.
To explore the effects of the dilution factor on our distance and explosion epoch, we perform the calculations with three different dilution factors from \citet{1996ApJ...466..911E}, \citet{2005AA...439..671D} and \citet{2019AA...621A..29V}.
The photospheric velocity was estimated using the tight relation (i.e., Equation 1 of \citet{2012MNRAS.419.2783T}) between $v_{\rm Fe II 5169}$ and photospheric velocity. We extrapolated the velocities to the epochs of the photometric measurements using Equation 3 of \citet{2012MNRAS.419.2783T}.
The explosion epoch $t_0$ and the distance $D$ can be estimated simultaneously by fitting a line defined by $t = (\theta/{v_{\rm phot}})D + t_0$. 
Since EPM can only be applied at very early times, 
only the quantities (i.e. epochs, angular radius, temperature, dilution factor, and photospheric velocities) from less than 10 days after the explosion were used in the fit (see Figure \ref{fig:epm} and Table \ref{tab:para_epm}).
The results for these three models of the dilution factors (\citet{1996ApJ...466..911E}, \citet{2005AA...439..671D} and \citet{2019AA...621A..29V}) are $t_{0}=59918.79, D=16.61\,{\rm Mpc}$; $t_{0}=59918.21, D=21.78\,{\rm Mpc}$ and $t_{0}=59917.89, D=20.69\,{\rm Mpc}$, respectively.

Based on the empirically observed luminosity-velocity correlation, which indicates that more luminous SNe II have higher photospheric expansion velocities, the distance of SN II can be derived from the apparent magnitude and the envelope velocity (see Equation 1 of \citealt{2002ApJ...566L..63H}).
The apparent V-band magnitude (see Section \ref{sec:lightcurve}) and $v_{\rm Fe II 5169}$ (see Section \ref{sec:expansion velocities}) at 50 days of SN 2022acko were interpolated using Gaussian Process Regression and a power-law fit, respectively.
Using the apparent V-band magnitude ($m_V^{50}=16.42\pm 0.01$ mag) and Fe~{\sc ii} velocity ($v_{\rm FeII 5169}^{50}=2756\pm 210 \rm \ km\ s^{-1}$) at 50 days after the explosion, the SCM distance of SN 2022acko is estimated to be 21.97$\pm$1.98\,Mpc, adopting $\rm H_{0}=70 \rm km^{-1}\ Mpc^{-1}$. 
Adopting the calibration from \citep{2015AA...580L..15P}, \cite{2023MNRAS.524.2186V} arrived at an SCM distance of 23.4$\pm$3.9\, Mpc.
The distance of SN 2022acko estimated by each method are listed in Table \ref{tab:distance}.
We adopt the average of these distances (i.e., 19.8$\pm$2.8\,Mpc) as the final distance for our analysis.
This distance is slightly larger ($\sim \rm 0.8\,Mpc/0.1 mag$) than that used in \citet{2023ApJ...953L..18B} and \citet{2023MNRAS.524.2186V}.

\subsection{Metallicity}
To estimate the metallicity at the SN location, we extract the strong host emission line from the $t=6$\,d
spectrum from \citep{2023ApJ...953L..18B} to measure the flux ratio of host N~{\sc ii} 6583 and H$\alpha$ emission lines (i.e., The $\rm N_2$ indicator of \citealt{2004MNRAS.348L..59P}). We found
that $\rm N_{2}$ is -0.158 and 12 + log(O/H) = $\sim$8.81 (1.3$\rm Z_{\odot}$, assuming a solar abundance of 8.69 \citep{2009ARA&A..47..481A})\footnote{we note that $\rm N_{2}$ is beyond the valid area, however, from the Figure 1 of \citealt{2004MNRAS.348L..59P}, it is reasonable to assume the linear relation between oxygen abundance and $\rm N_{2}$ indicator could be extrapolate beyond the valid region}. 
A spectrum of host nucleus taken by 6dF galaxy survey is available in NED, from which, the $\rm N_2$ indicator of nucleus area is estimated as -0.0239, indicating a very high metallicity at the host center. Assuming an average metallicity gradient, $-0.47R_{25}^{-1}$ \citep{2004A&A...425..849P}, the metallicity at the SN location is 8.72, again, suggesting a higher metallicity environment of SN2022acko than solar.

The metallicity at the SN location for other low-luminosity SNe was also estimated in this work. The host emission lines of H$\alpha$ and N~{\sc ii} 6583 are extracted from SN spectra whenever possible. Otherwise, the metallicity of the host nucleus and the average metallicity gradient were combined to estimate the metallicity at the SN location. 
The metallicities of some SNe have already been calculated using the same procedure as described above; therefore, we adopted these values from the literature when available.
The metallicity of each object is listed in Table \ref{tab:m_llsne}. 
From these 17 LLSNe, we found that the metallicity distribution of the LLSNe environment is uniformly spread, with an average metallicity of  8.6$\pm$0.2. This suggests that the metallicity of SN 2022acko lies above average

\subsection{Follow-up Photometry and Spectroscopy}
After the discovery of SN 2022acko, follow-up optical photometric observations in Johnson BV and Sloan $gri$ bands started quickly using LJT with Yunnan faint object spectrograph and camera (YFOSC; \citealp{2019RAA....19..149W}). All CCD images were pre-processed following standard routines, including bias subtraction, flat-field correction, and cosmic ray removal.
Point-spread-function (PSF) photometry was performed for both the SN and the reference stars using PyRAF. 
For the $gri$-bands, pre-SN templates from Panoramic Survey Telescope \& Rapid Response System (Pan-STARRS) were used and for the BV bands, only PSF photometry was performed, as we have no pre-SN template. 
BV-bands Magnitudes of the reference stars are obtained from the AAVSO Photometric All-Sky Survey\footnote{https://www.aavso.org/} (APASS) and magnitudes in $gri$-bands are obtained from the Pan-STARRS \citep{2016arXiv161205560C}.

SN 2022acko was also observed in the $uvw2$, $uvm2$, $uvw1$, $u$, $b$, and $v$ bands using the Ultraviolet/Optical Telescope (UVOT) onboard the
Neil Gehrels Swift Observatory (\cite{2004ApJ...611.1005G,2005SSRv..120...95R}). Data reduction followed that of the Swift Optical Ultraviolet
Supernova Archive \citep{2014Ap&SS.354...89B}.
A 5 arcsec aperture was used in the co-added image for the photometry.
Photometric results of the reference stars and SN 2022acko are listed in 
Table \ref{tab:standardstar}, Table \ref{tab:opt_phot} and Table \ref{tab:swift}.
We made comparisons with light curves from \cite{2023ApJ...953L..18B} and found consistency in both datasets. We include complementary data from \cite{2023ApJ...953L..18B} in our analysis when necessary.

After the discovery of SN 2022acko, we also initiated a high-cadence optical spectroscopic follow-up campaign using LJT.
The spectra covered the phases from $\sim$1.5\,d to $\sim$61\,d after the explosion. Details of the spectroscopic observations are listed in Table \ref{tab:spectra}.
All spectra were reduced using standard IRAF pipelines. The procedures included bias and flat-field corrections, cosmic ray removal, wavelength calibration, and flux calibration. The spectra were corrected for atmospheric extinction using the extinction curves of the local observatories. Telluric absorption was removed through comparison with the standard star spectrum.

\section{PHOTOMETRIC EVOLUTION}
\subsection{Lightcurve}
\label{sec:lightcurve}
\begin{figure}
	\includegraphics[width=1\columnwidth]{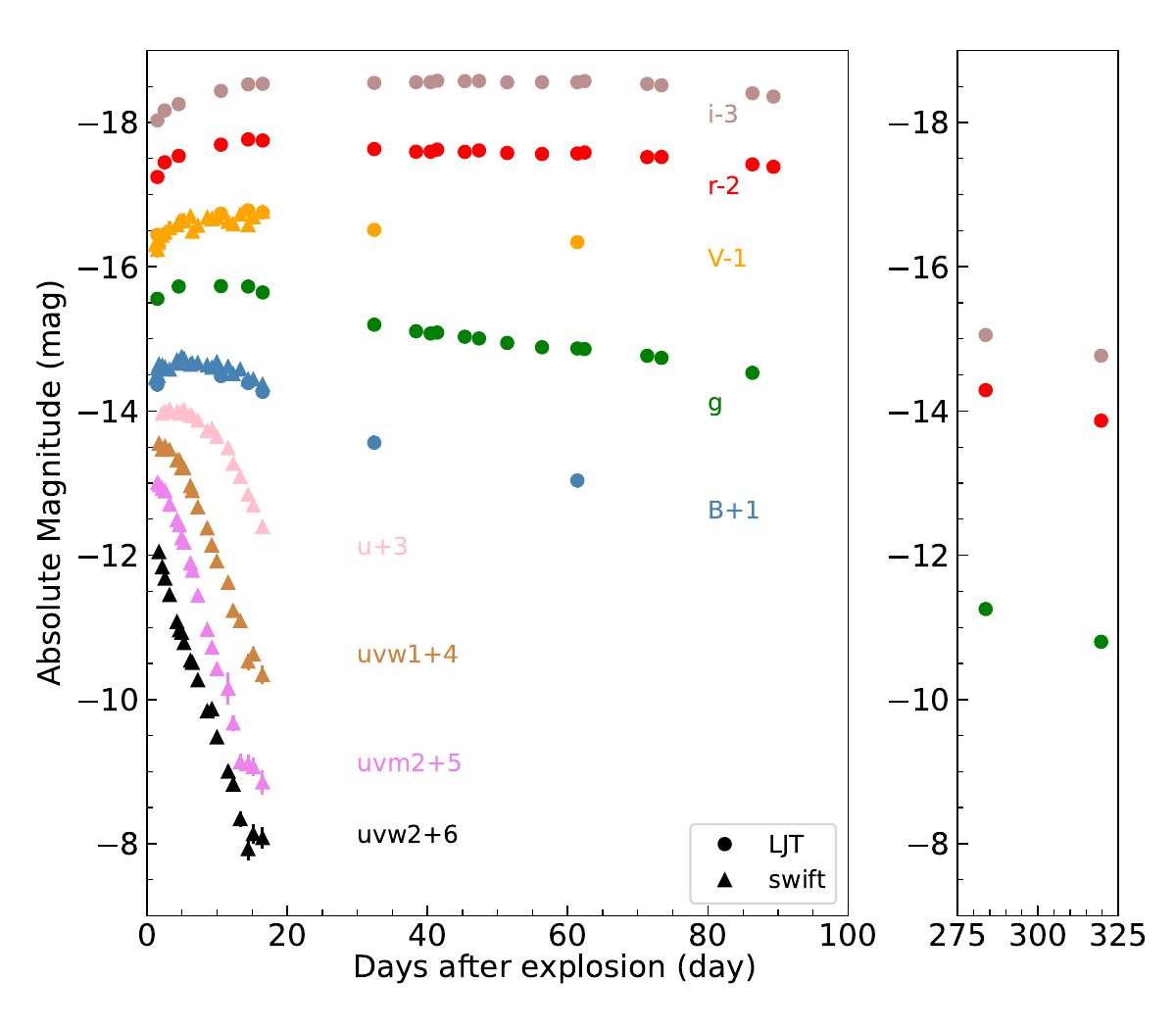}
	\caption{The UV and optical light curves of SN 2022acko. The light curves of different bands have been vertically shifted for clarity. }
	\label{fig:lc}
\end{figure}

\begin{figure*}
	\includegraphics[width=2\columnwidth]{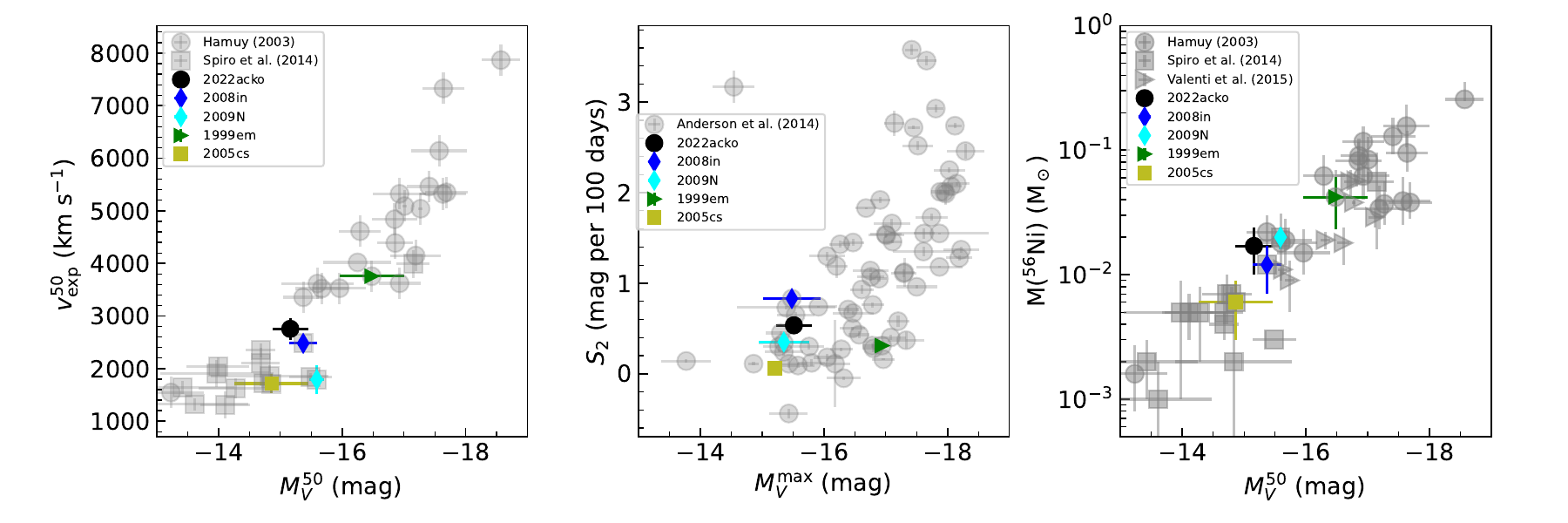}
	\caption{The position of SN 2022acko in the SN II family in terms of various photometric and spectroscopic indicators, including the V-band absolute magnitude ($M^{50}_V$), expansion velocity of Fe~{\sc ii} $\lambda$5169 measured at 50\,d after the explosion  ($v_{\rm exp}^{50}$), the $^{56}$Ni mass, the maximum V-band absolute magnitude($M_{V}^{\rm max}$) and the decline rate of the plateau ($s_2$). 
		The SNe II sample in \citet{2003ApJ...582..905H,2014ApJ...786...67A,2014MNRAS.439.2873S,2015MNRAS.448.2608V} together with SN 2008in, SN 2009N, SN 2005cs and SN 1999em are shown for comparison.
 	}
	\label{fig:phot_loc}
\end{figure*}

Figure \ref{fig:lc} shows the UV and optical light curves of SN 2022acko, with the right panel highlighting the late-time $gri$-bands photometry. SN 2022acko achieves the V-band peak about seven days after the explosion, reaching a peak V-band magnitude of $M_V= -15.5 \pm 0.3$ mag, consistent with the results in \citet{2023ApJ...953L..18B}. 
The $s_2$ parameter, defined as the decline rate of the second, shallower slope in the V-band light curve, is $s_2 = 0.54$\,mag per 100 days. The V-band absolute magnitude at 50 days after the explosion ($M_V^{50}$) is $-15.2\pm 0.3$ mag. 
The decline rates of the $gri$ bands at the nebular phase (from $\sim$284\,d to $\sim$320\,d) are 0.013, 0.012 and 0.008 $\rm mag \ day^{-1}$, respectively.
To describe where SN2022acko stands throughout the Type II sample,
we compared the V-band absolute peak magnitude and $s_2$
parameter of SN 2022acko with those of the SN II sample \citep{2003ApJ...582..905H,2014ApJ...786...67A,2014MNRAS.439.2873S,2015MNRAS.448.2608V}, as well as the normal SN IIP 1999em, the classical LLSN 2005cs and two other LLSNe (2008in and 2009N) that have the most similar photometric properties to SN 2022acko. As shown in Figure \ref{fig:phot_loc}, SN 2022acko is located at the low-luminosity and slow-declining side of the entire SNe II sample.
The properties during the plateau phase (i.e., $M_V^{50}$ and $V_{\rm exp}^{50}$) are intermediate, bridging the gap between low-luminosity SNe II and normal SNe II.

\begin{figure}
	\centering  
	\subfigure{
		\includegraphics[width=0.75\linewidth]{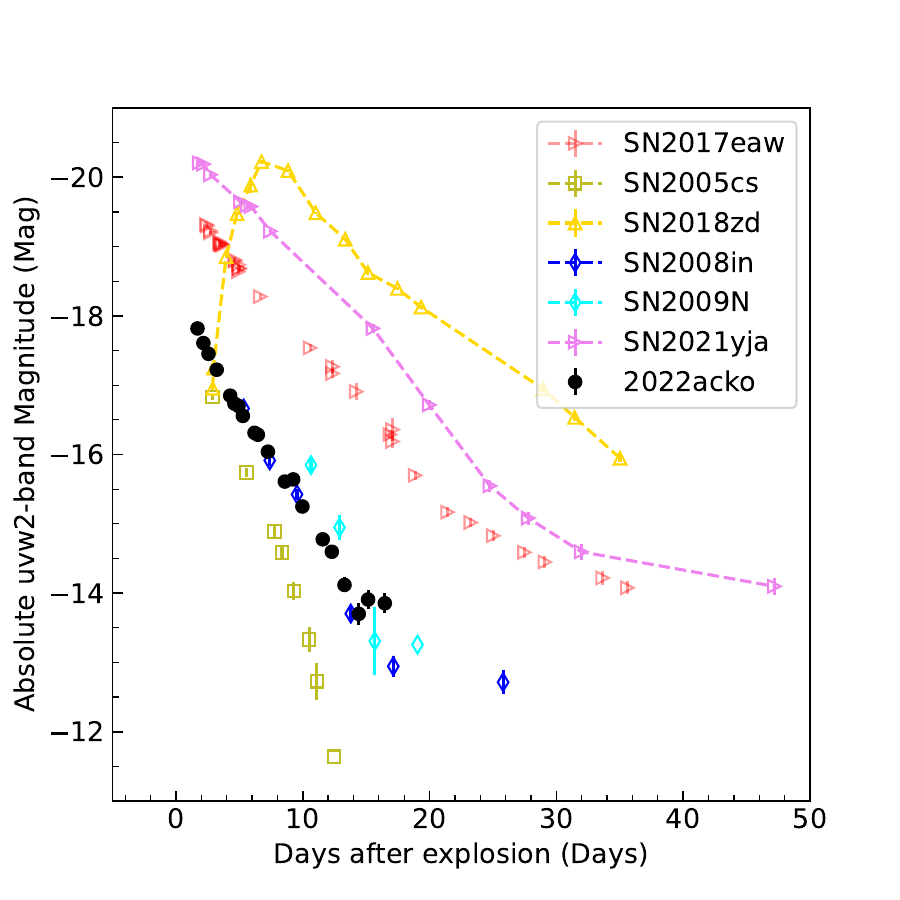}
	}
	\subfigure{
		\includegraphics[width=0.75\linewidth]{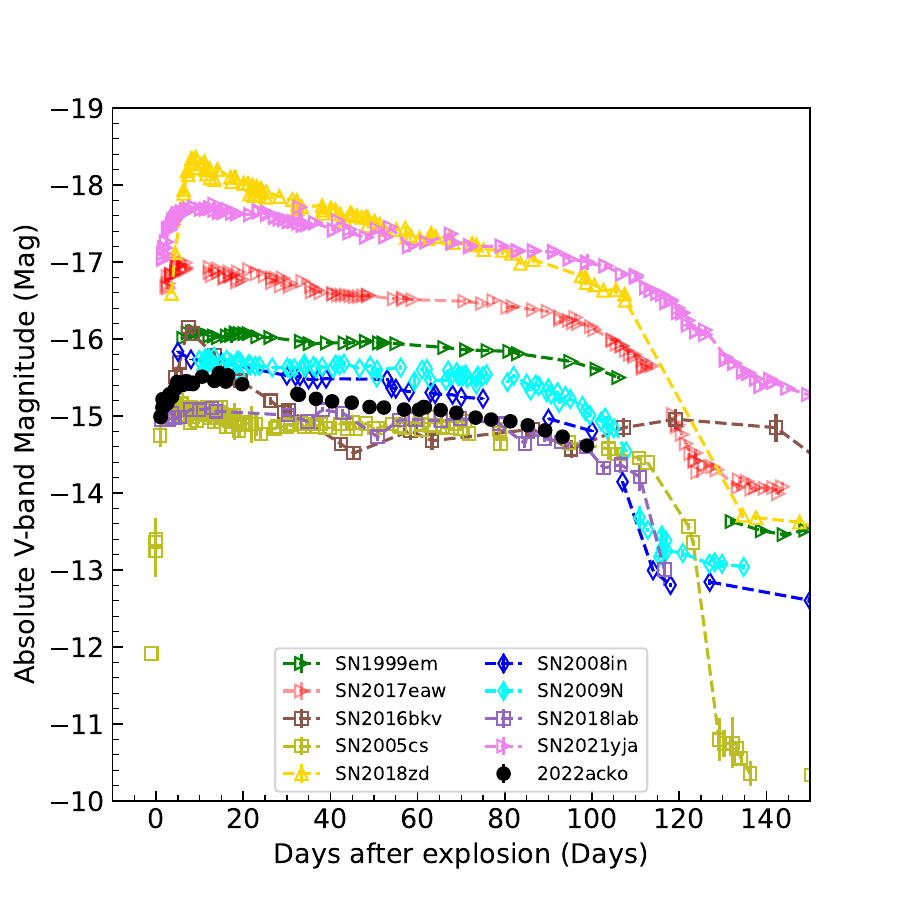}
	}
	\subfigure{
		\includegraphics[width=0.75\linewidth]{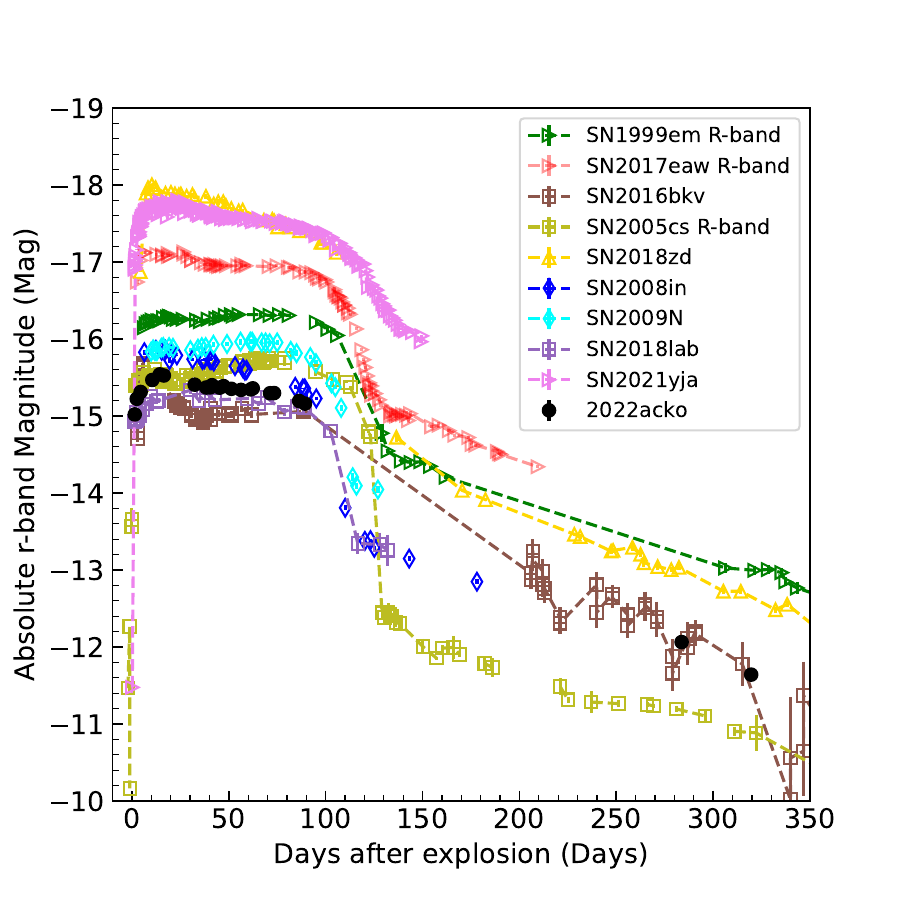}
	}
	
	\caption{$uvw2$-band (top-panel), V-band (mid-panel) and $r$-band (bottom-panel) light-curve comparisons between SN 2022acko and some well-studied SNe II, including the typical SN IIP 1999em \citep{2001ApJ...558..615H,2002PASP..114...35L,2003MNRAS.338..939E} and 2017eaw \citep{2019ApJ...875..136V,2019ApJ...876...19S},  the low-luminosity SN 2005cs \citep{2006MNRAS.370.1752P,2009MNRAS.394.2266P}, SN 2016bkv \citep{2018ApJ...859...78N,2018ApJ...861...63H}, SN 2018lab \citep{2023ApJ...945..107P}, SN 2008in \citep{2011ApJ...736...76R} and SN 2009N \citep{2011ApJ...736...76R}, the FI object SN 2018zd \citep{2020MNRAS.498...84Z}
		and the ledge object SN 2021yja \citep{2022ApJ...935...31H}
	}. 
	\label{fig:multibandcompare}
\end{figure}

To further describe the photometric properties of SN 2022acko, we compared the $uvw2$, V, and $r$ bands light curves with other well-studied objects in Figure \ref{fig:multibandcompare}.
In the $uvw2$-band, the magnitude of SN 2022acko peaks at $\sim\,-18.0$ mag and shows a monotonic decline starting $\sim$ 1.7 days after the explosion. 
Its $uvw2$-band magnitude is brighter than that of SN 2005cs and is closest to that of SN 2008in.
Compared with the sharp rise that is shown in low luminosity SN 2005cs and SN 2018lab, the V-band light curve of SN2022acko exhibits a smoother and longer rise before reaching the maximum. It does not show an early bump like SN 2018zd and SN 2016bkv as well.  
Different from those SNe IIP that show an increase in R/r-band after some time after peak (e.g., SN 2005cs, SN 1999em, SN 2009N, and SN 2016bkv), 
the r-band light curve of SN 2022acko shows a monotonic decline after peak. At the nebular phase, the r-band magnitude is similar to that of SN 2016bkv, suggesting a similar nickel mass synthesized in the explosion.

\begin{figure}
	\centering  
	\subfigure{
		\includegraphics[width=0.75\linewidth]{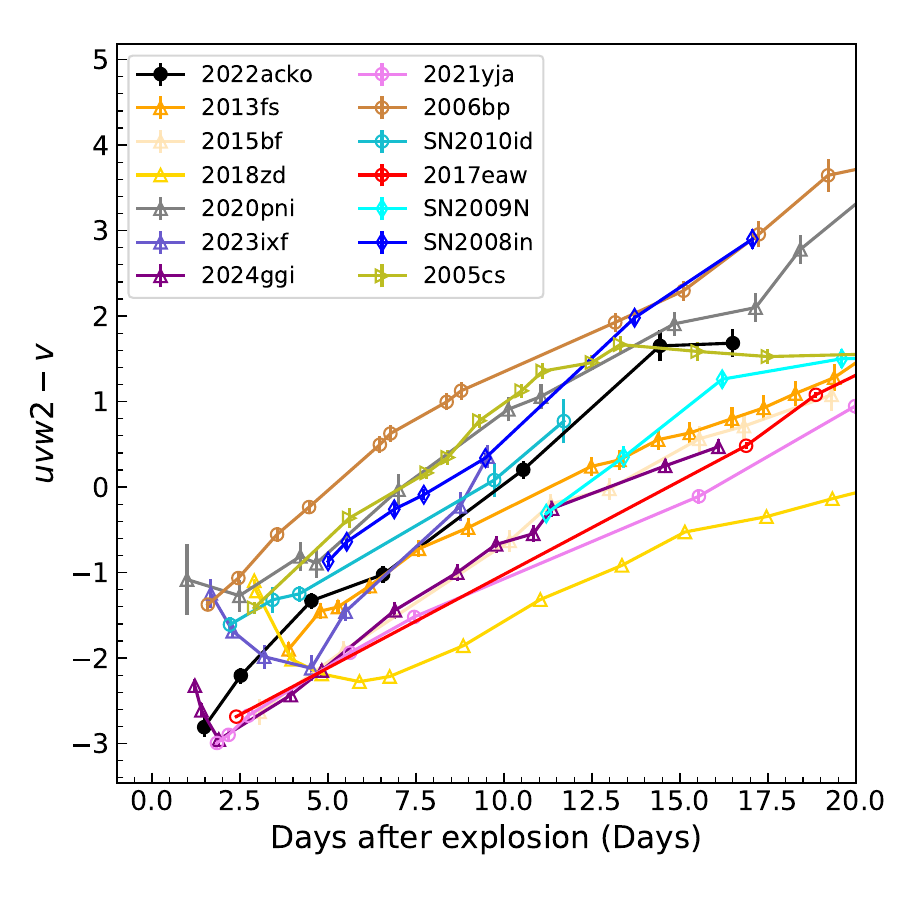}
	}
	\subfigure{
		\includegraphics[width=0.75\linewidth]{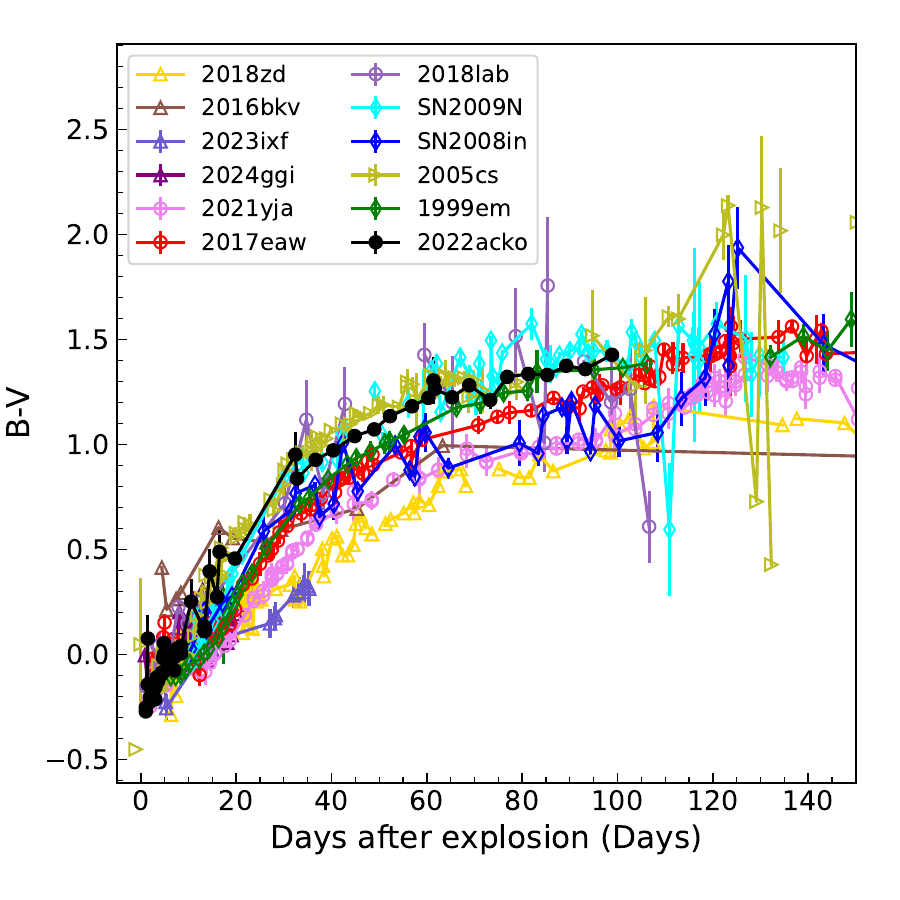}
	}
	\caption{The $uvw2-v$ and B-V color curve evolution of SN 2022acko, along with that of other well-studied SNe II. All the colours have been corrected for both the Galactic and host-galaxy reddening.} 
	\label{fig:colorcurve}
\end{figure}

\subsection{Color curve}
In Figure \ref{fig:colorcurve}, the $uvw2-v$ and B-V color of SN 2022acko was compared with that of a sample of SNe II, mainly including the individuals that show prominent flash-ionized emission lines (SN 2013fs, SN 2015bf, SN 2018zd, SN 2020pni, SN 2023ixf \citep{2023ApJ...954L..42J,2024Natur.627..759Z} and SN 2024ggi \citep{2024ApJ...972L..15S}), the SNe II that show broad ``ledge'' emission at $\sim 4600$ \AA \ (SN 2006bp, SN 2010id, SN 2017eaw, and SN 2021yja), and the low-luminosity object (SN 2005cs, SN 2008in and SN 2009N). For the flash objects, the $uvw2-v$ colors show a pronounced blueward evolution at the earliest phase, see SN2018zd, SN 2020pni, SN 2023ixf and SN 2024ggi in Figure \ref{fig:colorcurve}, which \cite{2021NatAs...5..903H} proposed that it is a signature of the shock breakout of a dense CSM.
\citet{2024ApJ...972L..15S} likewise attribute the initial blueward color evolution to interaction with dense CSM.
However, we do not observe such a turnaround in SN 2022acko and other ``ledge" objects; the $uvw2-v$ color is blue in early phases and evolves toward red for the whole time. 
\citet{2024ApJ...972L..15S} found that the turnover in color evolution seems to occur near the time that flash-ionized features disappear. Therefore, the lack of blueward evolution in SN 2022acko and other ``ledge" objects may indicate a much more confined CSM compared with those strong and long-lasting FI objects.
Similar to the color evolution of other SNe II, the B-V color evolves toward red as the large envelope of the RSG progenitor expands and cools. 
SN 2022acko is redder at the photospheric phase, suggesting a relatively lower temperature.

\subsection{Bolometric lightcurve}
We constructed the SED to estimate the bolometric luminosity by fitting a blackbody model to the available multiband photometry.  When $Swift$ UV photometry was available (i.e., $< 20$\,days after the explosion), these UV data were included in the fitting process. In the later phases, only optical band data were considered. 
At early and plateau phases, the bolometric luminosity of SN 2022acko was calculated by integrating the SED from 1600\AA\,to 136000 \AA. 
At nebular phases, since we only have $gri$-bands photometry, the bolometric correction (BC) in \cite{2014MNRAS.437.3848L} was used to transform the $g$-band light curve to the bolometric light curve. 
In order to evaluate the compatibility across luminosity measurements obtained at different stages and through other methods, we also applied the BC method to the early and plateau phases.
Unless otherwise specified, the bolometric luminosity in our work refers to the estimate obtained from blackbody integration. 
\citet{2023ApJ...953L..18B} found that a blackbody spectrum is a poor approximation for the UV flux, even as early as 5 days after the explosion. Therefore, we apply the BC of the shock cooling regime only for the first two epochs. At later epochs, the BC of radiatively/recombination regime was adopted. 
The primary discrepancy between these two BC approaches arises from the different treatment of UV flux, specifically, employing a blackbody (fitted from the $B$-to $K$- band) extrapolation to 0 \AA \ during the cooling phase and a linear extrapolation to zero flux at 2000 \AA \ for epochs of no strong cooling.

Since the blackbody integration approach in our work adopted a different treatment of UV flux and the integration wavelength range with that of \citet{2014MNRAS.437.3848L}, discrepancies in luminosity between the two methods are expected.
At the early phases when the SED peaks in the UV, the lack of UV-band photometry in the fitting procedure of \citet{2014MNRAS.437.3848L} may lead to a lower temperature than that derived by blackbody fitting in our work, resulting in a substantial discrepancy in luminosity. Specifically, the luminosity calculated using the bolometric correction (BC) method is approximately half (~50\%) of that obtained through blackbody integration at phases earlier than five days after the explosion. 
As the SED peak shifts to the optical, the contribution of UV flux to the overall luminosity diminishes, reducing the luminosity difference between the two methodologies. For example, at plateau phases in the case of SN 2022acko, the luminosity ratio is $L_{\rm Lyman}/L_{\rm BB} \sim 0.87$.

By integrating the flux redward of a given
wavelength from the combined UV and optical spectra, \citet{2023ApJ...953L..18B} found that only 53\% of the flux is captured at day 5 when the bluest filter is the B band, but this fraction 
increases to 95\% by day 19. Similarly, we estimated the UV flux (integrated from 1600\AA \  to 3000\AA \ ) ratio, finding that the UV flux contributes 73\%, 61\%, and 43\% of the total luminosity at +1.48\,d, +2.52\,d and +4.53\,d after the explosion, respectively. At 16.5 days after the explosion, the fraction of UV flux decreases to 5\%.

\begin{figure}
    \centering  

	\subfigure{
		\includegraphics[width=0.9\linewidth]{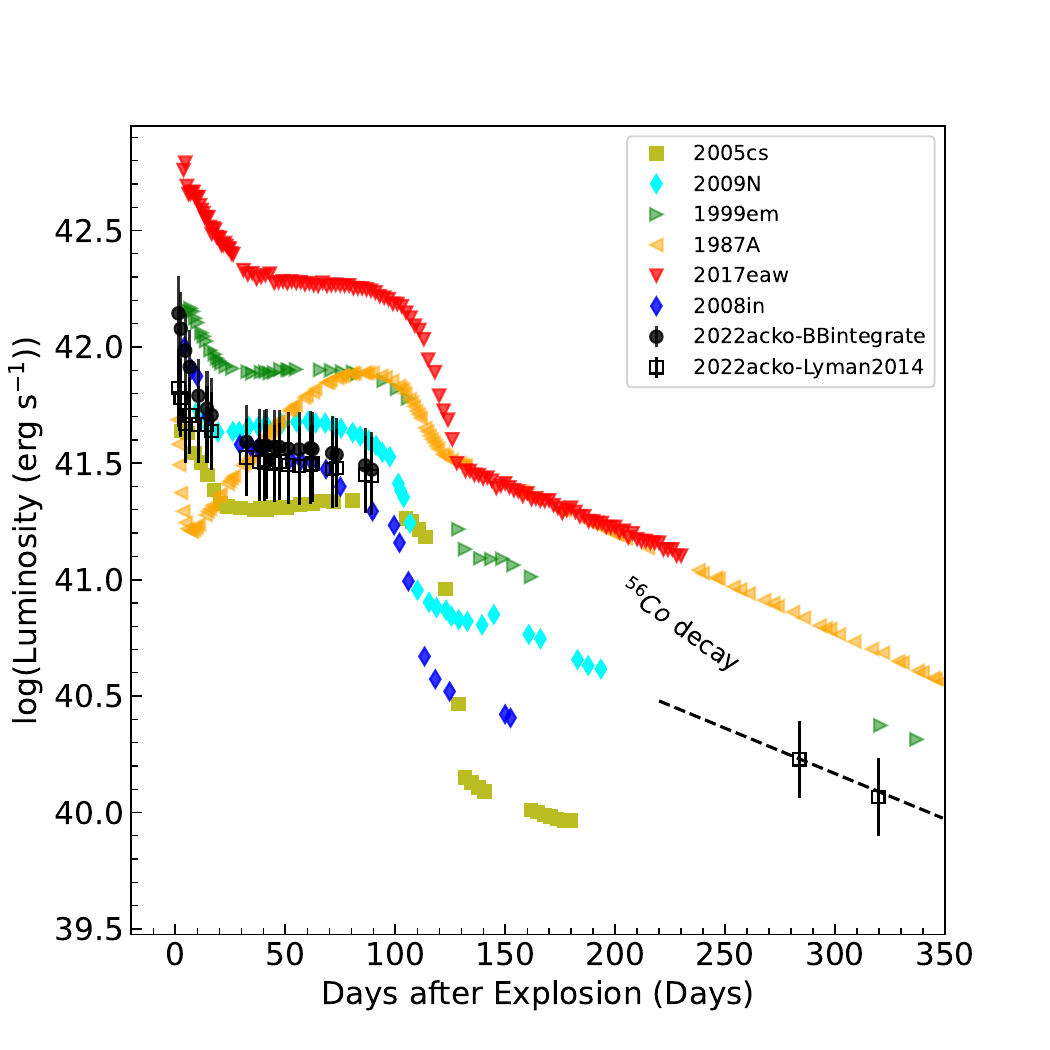}
	}
	\subfigure{
		\includegraphics[width=0.9\linewidth]{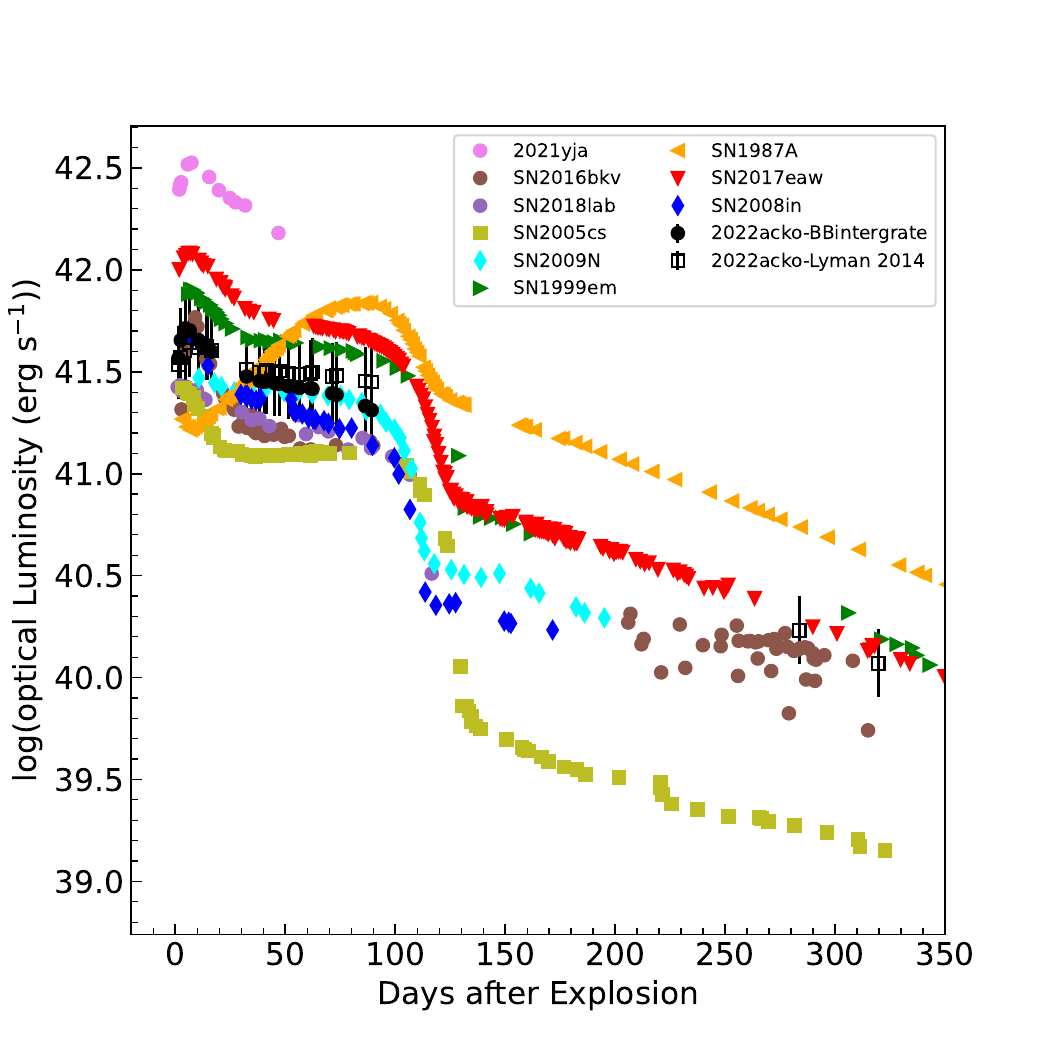}
	}
	\caption{Pseudo-bolometric light curve (top panel) and the pseudo-bolometric light curve in the optical band 
    (bottom panel) of SN 2022acko, compared with that of a few well-studied SNe II. SED integration and bolometric corrections were adopted to estimate the luminosity of SN 2022acko. The SED was constructed by fitting a blackbody to the photometry data. We integrate the blackbody spectrum from 1600\AA \ to 136000 \AA \, to estimate the bolometric luminosity (upper panel) and from 3000 \AA \ to 11800 \AA \, to estimate optical luminosity (lower panel)). 
    } 
	\label{fig:bolo}
\end{figure}

The bolometric light curve of SN 2022acko and the comparison with other well-studied samples are shown in Figure \ref{fig:bolo}. The bolometric luminosity of SN 2022acko reaches a luminosity of $\rm 1.4\times10^{42}\ erg s^{-1}$ immediately after the explosion, following a rapid decline until the bolometric luminosity entering into the plateau phase at about $\sim$ 25 days after the explosion. 
The plateau luminosity ( $\rm 3.7\times10^{41}\ erg \ s^{-1}$) is the most similar to that of SN 2008in.
The optical bolometric luminosity (which was estimated by integrating the SED from 3000\AA \  to 11800 \AA \ ) was compared with that of the same sample above, a few LLSNe, in addition to that of SN 2021yja, a "ledge" object but with high luminosity.
After a rise time of $\sim 5$ days, the optical bolometric luminosity peaks with a luminosity of $\rm 4.8\times10^{41}\ erg \ s^{-1}$. 
At the epoch of the optical peak, the SED peaks in UV. The B-V color is $-0.09\pm 0.01$\,mag and the SED is still blue. At $\sim 10$ days after explosion, the SED peak enters the
optical. At this time, the B-V color is $0.25 \pm 0.1$mag.
The optical bolometric light curve of SN 2022acko shows a clear rise at early epochs. 
Such an early rise is also shown in SN 2021yja, SN 2017eaw, and is the most prominent in SN 2016bkv. 
Compared with the plateau luminosity, the very strong peak around 7 days after the explosion in the flash object SN 2016bkv should be powered by a combination of shock cooling and circumstellar interaction \citep{2018ApJ...861...63H, 2018ApJ...859...78N}.
We noticed that the early peak of SN 2022acko is not that sharp compared with that of SN 2016bkv, possibly suggesting that the early CSI in SN 2022acko is not as intense as that of SN 2016bkv.

\section{SPECTROSCOPIC ANALYSIS}

\subsection{Evolution}
\begin{figure*}
	\includegraphics[width=1.9\columnwidth]{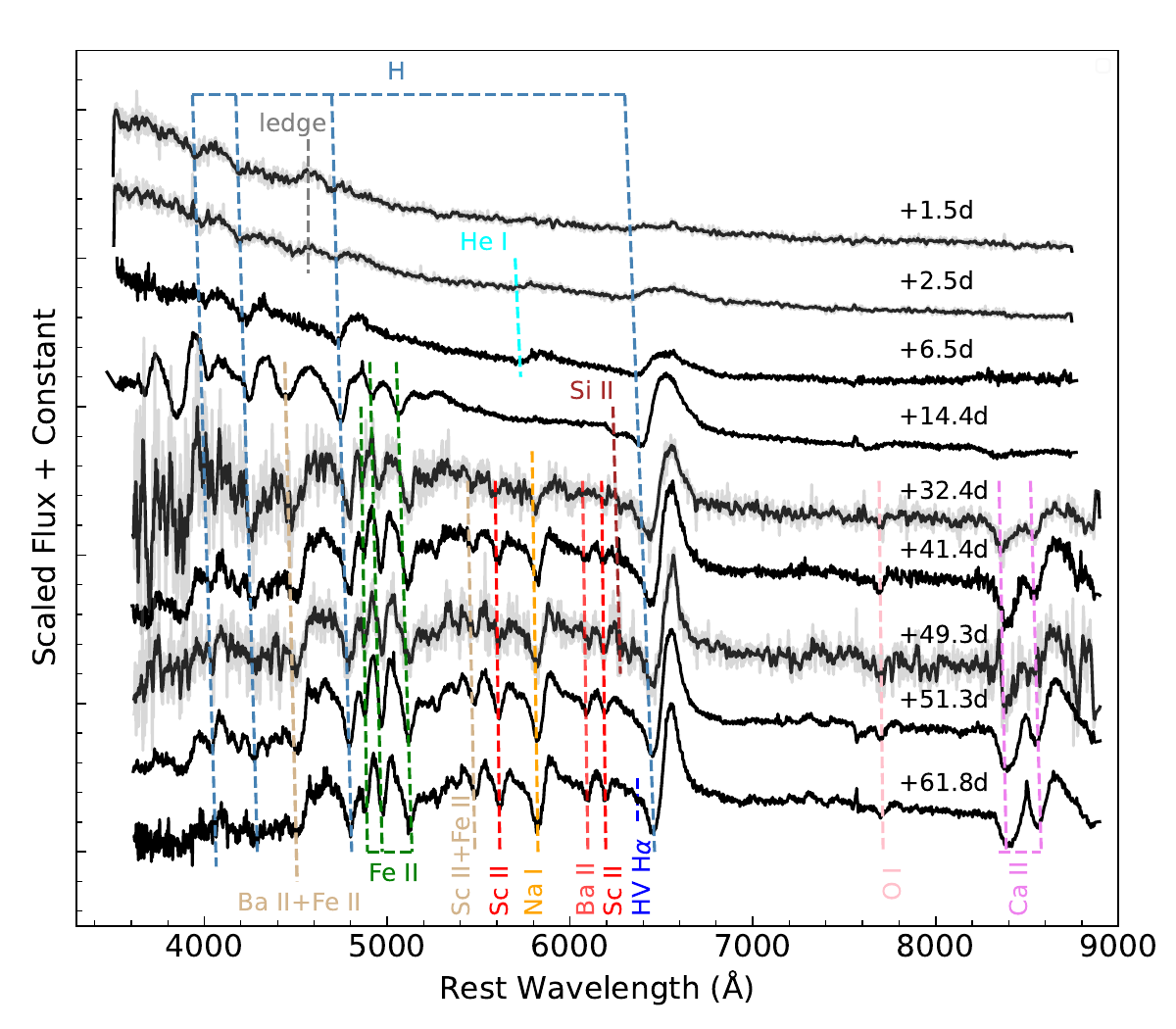}
 \centering

	\caption{The optical spectra of SN 2022acko, shifted vertically for clarity. The numbers on the right side mark the epochs of the spectra in days after the explosion. Spectral lines of different species are marked by dashed lines with different colors. The broad emission at 4600 \AA \ in the $t=1.5$\,d and $t=2.5$\,d spectra was marked by gray dashed line. }
	\label{fig:spec_show}
\end{figure*}

Figure \ref{fig:spec_show} illustrates the spectral evolution of SN 2022acko, spanning from $t$ = 1.5\,d to $t$ = 62\,d after the estimated explosion date. 
The broad ``ledge'' emission feature peaked near 4600 \AA\, has been detected in the $t$=1.5\,d and $t$=2.5\,d spectra of this SN. Concurrently, the $\rm H\alpha$ show broad P-Cgyni profiles with very shallow absorption. The absorption minimum is located at $\sim$12000 $\rm km \ s^{-1}$ at $t$=1.5\,d. Moreover, a weak narrow $\rm H\alpha$ emission line was superposed on the broad emission component of $\rm H\alpha$ at $t$=1.5\,d, see Figure \ref{fig:ha_emi}. 
To measure the FWHM of this narrow component of $\rm H\alpha$ at $t$=1.5\,d, we first subtract the underlying continuum and fit with a Gaussian to the remaining narrow $\rm H\alpha$ component, as shown in Figure \ref{fig:ha_emi}. The O~{\sc i} 5577 sky emission line was also fitted with a gaussian function to estimate the instrumental resolution.
The FWHM of the narrow $\rm H\alpha$ is about $1390 \pm 400\,\rm km\ s^{-1}$,  which is larger than the instrumental resolution of 840\,$\rm km\ s^{-1}$. After the instrumental correction determined by $\rm FWHM_{cor} = (FWHM_{obs}^{2}-FWHM_{inst}^{2})^{1/2}$), the FWHM of the narrow $\rm H\alpha$ is $\rm \sim 1100 \pm 550 \ km\ s^{-1}$.

\begin{figure*}
	\centering  
	\subfigure{
		\includegraphics[width=1.5\columnwidth]{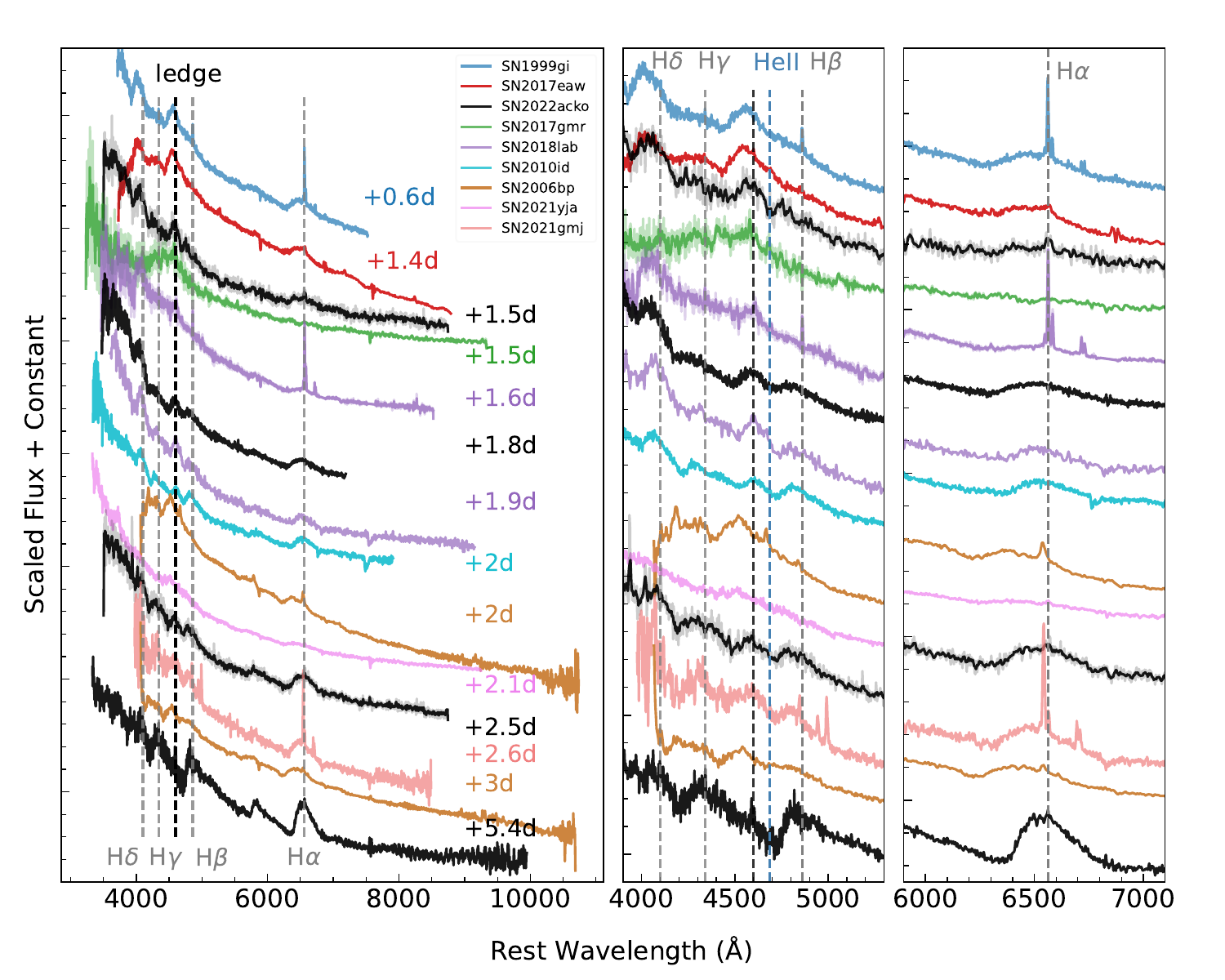}
	}

	\subfigure{
		\includegraphics[width=1.5\columnwidth]{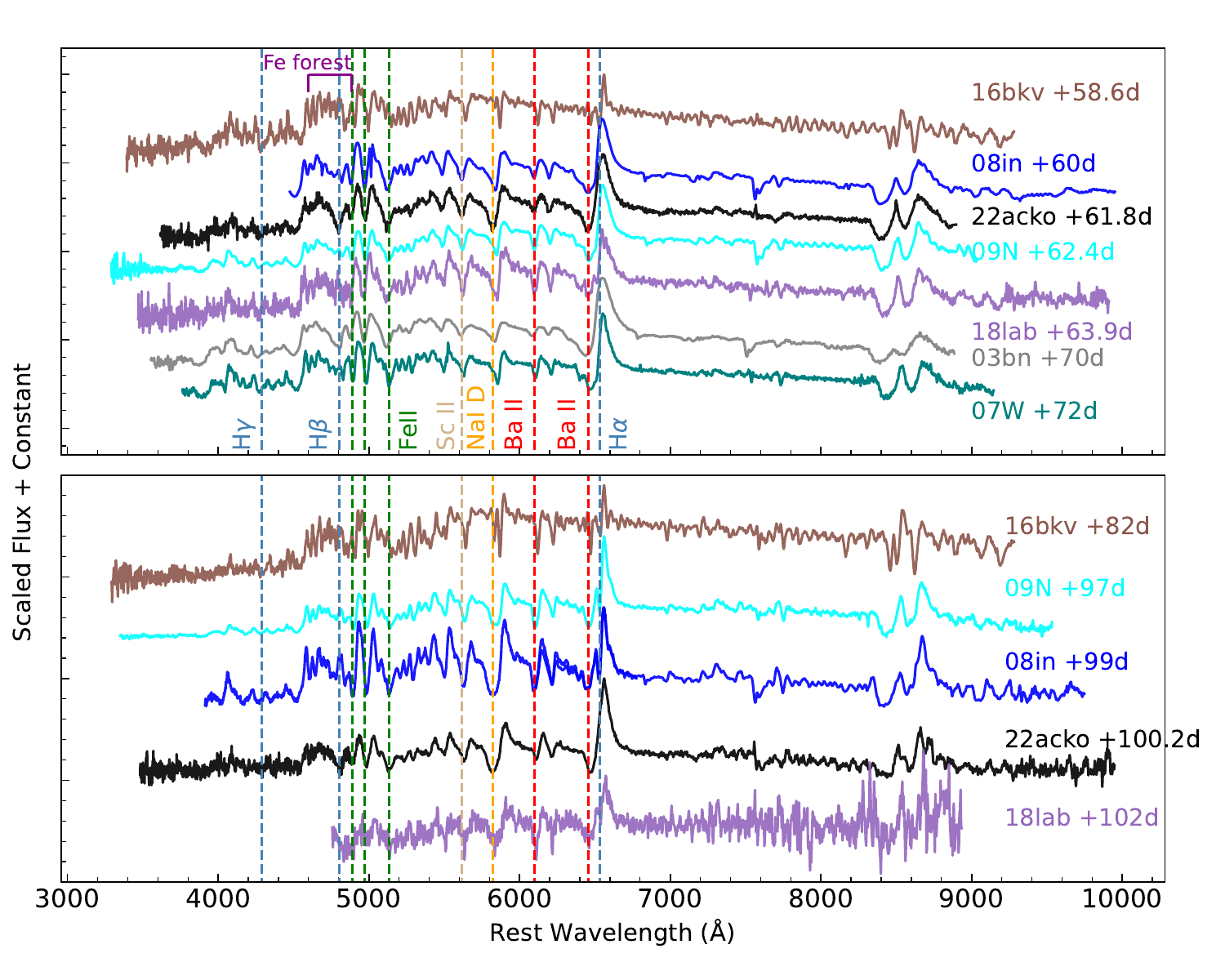}
	}

	\caption{Spectral comparisons of SN 2022acko at early (upper panel) and photospheric phase (lower panel). The early spectra of SN 2022acko were compared with the objects that show a similar ``ledge'' feature (represented by the black dashed line in the upper panel), 
	including SN 2010id \citep{2011ApJ...736..159G}, SN 2006bp \citep{2007ApJ...666.1093Q}, SN 2017eaw \citep{2019ApJ...875..136V}, SN 1999gi \citep{2002AJ....124.2490L}, SN 2018lab \citep{2023ApJ...945..107P}, SN 2021yja \citep{2022ApJ...935...31H}, SN 2017gmr \citep{2019ApJ...885...43A}, and SN 2021gmj \citep{2024MNRAS.528.4209M}.
    The $t=1.8$d spectrum of SN 2022acko which was taken from \citet{2023ApJ...953L..18B} was also shown in the upper panel. 
    The spectra of SN 2022acko at photospheric phase ($\rm \sim 60 \ days$ and $\rm \sim 100\ $days after the explosion) were compared with SN 2003bn \citep{2017ApJ...850...89G}, SN 2007W \citep{2017ApJ...850...89G} and the Low-luminosity SN 2008in, SN 2009N, SN 2018lab and SN 2016bkv.} 
	\label{fig:early_ps_compare}
\end{figure*}

The ``ledge'' feature has been detected in the spectra of several SNe soon after the explosion, e.g., SN 1999gi, SN 2006bp, SN 2017gmr, SN 2017eaw. \citet{2023ApJ...945..107P} noticed that four out of six low-luminosity objects with spectra taken within five days after explosion show such features. 
In the upper panel of Figure \ref{fig:early_ps_compare}, we compare the early spectra of SN 2022acko with those of other SNe that show ``ledge'' features. We noticed that the majority of these samples, except SN 2017gmr\footnote{Narrow H$\alpha$ emission, with a Gaussian FWHM velocity of $\sim 55 \rm km\ s^{-1}$, are detected at the $t=1.5$\,d spectrum of SN2017gmr, however, the broad P-Cgyni profile is not that evident.},
show broad H$\alpha$ P-Cgyni profiles with very shallow absorption. Additionally, in SN 2022acko, SN 2017eaw, SN 2021yja and SN 2006bp, a narrow H$\alpha$ emission line was superposed on the broad component at early phases. 
\cite{2007ApJ...666.1093Q} had previously identified the narrow H$\alpha$ and He~{\sc ii} 4686 emission lines in the spectra of SN 2006bp within 3 days after the explosion. 
Although SN 2017eaw was initially thought to have negligible or weak CSI at its early stages \citep{2019ApJ...875..136V,2024ApJ...970..189J}, \cite{2019MNRAS.485.1990R} identified a narrow H$\alpha$ emission feature blueshifted by $\sim \rm 160 km \ s^{-1}$ in their $t=1.4$\,d spectrum. 
In SN 2021yja, we noticed a weak, narrow H$\alpha$ emission in the $t=2.1$\,d spectrum, see Figure 2 of \citet{2022ApJ...935...31H}.
Notably, for SN 2022acko, the FWHM of the narrow H$\alpha$, around $\rm 1100 \pm 550\ km\ s^{-1}$, is broader (though the lower limit is not that much broader) than that arises from the host galaxy H~{\sc ii} region.
Considering the low resolution and low signal-to-noise ratio of the $t=1.5$\,d spectrum, it is challenging to determine whether the narrow H$\alpha$ observed in the $t=1.5$\,d spectrum of SN 2022acko originates from the CSI or an H~{\sc ii} region. However, given the similar H$\alpha$ line morphology of SN 2022acko compared to that of SN 2006bp and SN 2017eaw, as well as the transition from FI profile to ledge profile in some standard FI object (see the fourth paragraph of Section \ref{Signs of circumstellar interaction}), we prefer the  interpretation that the narrow H$\alpha$ emissions in SN 2022acko arise from CSI.
In the case of SN 1999gi, SN 2018lab and SN 2021gmj, 
it is difficult to determine whether the CSI-associated narrow H$\alpha$ line exists since H$\alpha$ emission was contaminated by host galaxy H~{\sc ii} region. The spectrum of SN 2010id does not exhibit narrow H$\alpha$ lines, indicating the possibility that such lines may have been present earlier than observed or may not exist at all.
Interpretation of the ``ledge'' feature varies and we discuss it in Section 5.2.1.

After the ``ledge'' feature disappeared in our $t = 6.5$\,d spectra, the spectra of SN 2022acko evolved into a typical SN II. At $t = 6.5$\,d, only the P-Cgyni profile of hydrogen and helium were present. The He~{\sc i} feature vanished and was replaced by the Na~{\sc i} line at a similar position at $t = 14$\,d. Fe~{\sc ii} 5169, 5018 and Fe~{\sc ii} blends began to appear in the $t = 14$\,d spectra. Other metal lines, like Ba~{\sc ii}, Sc~{\sc ii}, and blends lines (e.g.Ba~{\sc ii}+Fe~{\sc ii}, Sc~{\sc ii}+Fe~{\sc ii}), as well as O~{\sc i} 7774 and Ca~{\sc ii} NIR triplet was evident in the $t = 32$\,d spectrum. 
In the spectra from 14 days to 49 days, a shallow absorption feature detected near 6240\AA \ is attributed to Si~{\sc ii} 6355. 
In the spectra of $t=61.8$\,d, a narrow notch appeared on the blueward side of H$\alpha$ (located near 6370 \AA  \footnote{This notch can not be Si~{\sc ii} 6355, since its absorption minima is larger than the rest wavelength of Si~{\sc ii} 6355.}) and persisted until $t= \sim $ 100\,d (see Figure 2 in \citealt{2023ApJ...953L..18B}). We interpret this as the high-velocity component of H$\alpha$, since its velocity is, on average, $\rm \sim 2500\ km\ s^{-1}$ larger than Fe~{\sc ii} 5169 if it was Ba~{\sc ii} 6497.

In the photospheric spectra after the mid-plateau phase of SN 2022acko, we found that the Ba~{\sc ii} 6497 line, typical in subluminous SNe, is absent. As shown in the lower panel of Figure \ref{fig:early_ps_compare}, Ba~{\sc ii} 6497 is detectable in SN 2009N and SN 2008in around $t\sim$ 60 days, but it remains undetectable in SN 2022acko, even up to $\sim$ 100 days after the explosion.
\cite{1998ApJ...498L.129T} suggested that Strong Ba~{\sc ii} lines are more likely a temperature effect rather than a result of Ba overabundance. Ba~{\sc ii} lines appear below $\sim$ 6000\,K \citep{1999ApJS..121..233H}. The temperature of SN 2022acko decreases to $\sim$ 6000\,K at around $\sim 35$ days. Ba~{\sc ii} 6142 can be detected in our $t=41$\,d spectrum and becomes stronger as the temperature decreases. However, Ba~{\sc ii} 6497 does not follow a similar evolutionary path. 

Besides, the ``Fe line forest", a series of Fe group absorption lines around H$\beta$ (approximately around 4600\AA \  to 4900\AA), see footnote 28 of \citet{2017ApJ...850...89G}, often seen in faint SNe with low ejecta temperatures and/or in high metallicity environments \citep{2017ApJ...850...89G}, such as SN 2008in and SN 2009N, is not detected in SN 2022acko.  
\citet{2023ApJ...953L..18B} conclude that there is a significant Fe absorption in the UV spectra. 
SN 2022acko is a low-luminosity object situated in a high-metallicity environment, which means that the ``Fe line forest" and Ba~{\sc ii} 6497 in SN 2022acko should be as obvious as that of other LLSNe samples. 
However, the characteristics of Ba~{\sc ii} 6497 and ``Fe line forest" in SN 2022acko bear more resemblance to those of standard SNe II, see the lower panel of Figure \ref{fig:early_ps_compare}. 
It is plausible that the contributions of Ba~{\sc ii} 6497 and H$\alpha$, as well as the ``Fe line forest" and H$\beta$, could blend into a singular spectral feature, potentially masking the presence of these lines.
Note that the profile of H$\alpha$ and H$\beta$ might show some visible distortion (e.g., increased line width) due to blending with nearby lines, though such distortions are not apparent in SN 2022acko.
\cite{2017ApJ...850...89G} suggested that SNe II lacking 
the Fe~{\sc ii} line forest distinctly differ from those exhibiting this feature.
Therefore, the non-detection of Ba~{\sc ii} 6497 and ``Fe line forest" in SN 2022acko may indicate underlying distinctions between this SN and other LLSNe\footnote{Such discussion is beyond the scope of our work.}.

\subsection{Expansion velocities}
\label{sec:expansion velocities}
\begin{figure}
	\includegraphics[width=0.9\columnwidth]{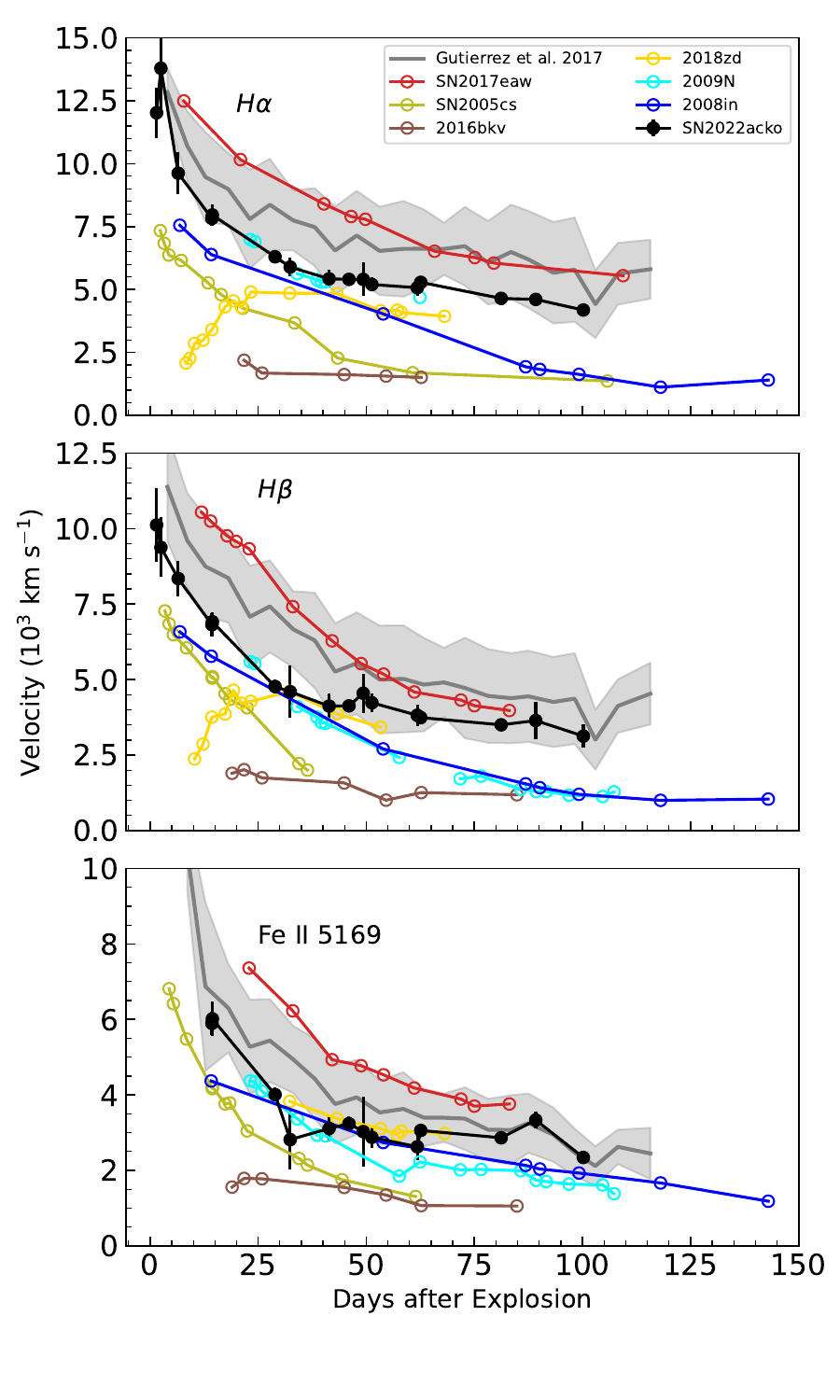}
        \centering
 	\caption{The velocity evolution of H$\alpha$, H$\beta$ and Fe~{\sc ii} $\lambda$ 5169 lines of SN 2022acko, along with comparison with some well-studied type II SNe and the average velocity evolution of SNe II sample of \citet{2017ApJ...850...89G}.}
	\label{fig:velocity_compare}
\end{figure}

The expansion velocities of Hydrogen and Fe lines were measured from the absorption minima of P-Cgyni profiles. Figure \ref{fig:velocity_compare} shows the velocities of H$\alpha$, H$\beta$, and Fe~{\sc ii} 5169, compared with other well-studied samples.
As shown in Figure \ref{fig:velocity_compare}, the velocities of SN 2022acko are generally 1$\sigma$ lower than the average value of the SN II sample in \cite{2017ApJ...850...89G}. 
SN 2022acko follows the positive luminosity-velocity correlation at the mid-plateau phase \citet{2003ApJ...582..905H}.
(i.e., $M_{V}^{50} - V_{\rm exp}^{50}$), placing it in an intermediate position that connects the low-luminosity and typical samples, see the first panel of Figure \ref{fig:phot_loc}.
The velocity evolution is the most similar to that of SN 2008in and SN 2009N before the mid-plateau phase. However, after the mid-plateau phase, the velocity of SN 2022acko declines more slowly, particularly in H$\beta$,  resulting in it being faster than SN 2009N and SN 2008in, despite similar velocities before the mid-plateau. We suggested that this higher expansion rate in velocity post-mid-plateau could be why 
Ba~{\sc ii} 6497 and the Fe line forest are not present in SN 2022acko; a higher expansion rate at the base of the H-rich envelope may cause them merged with the H$\alpha$ and H$\beta$ lines, respectively \citep{2017MNRAS.466...34L}.


\begin{figure}
	\includegraphics[width=1\columnwidth]{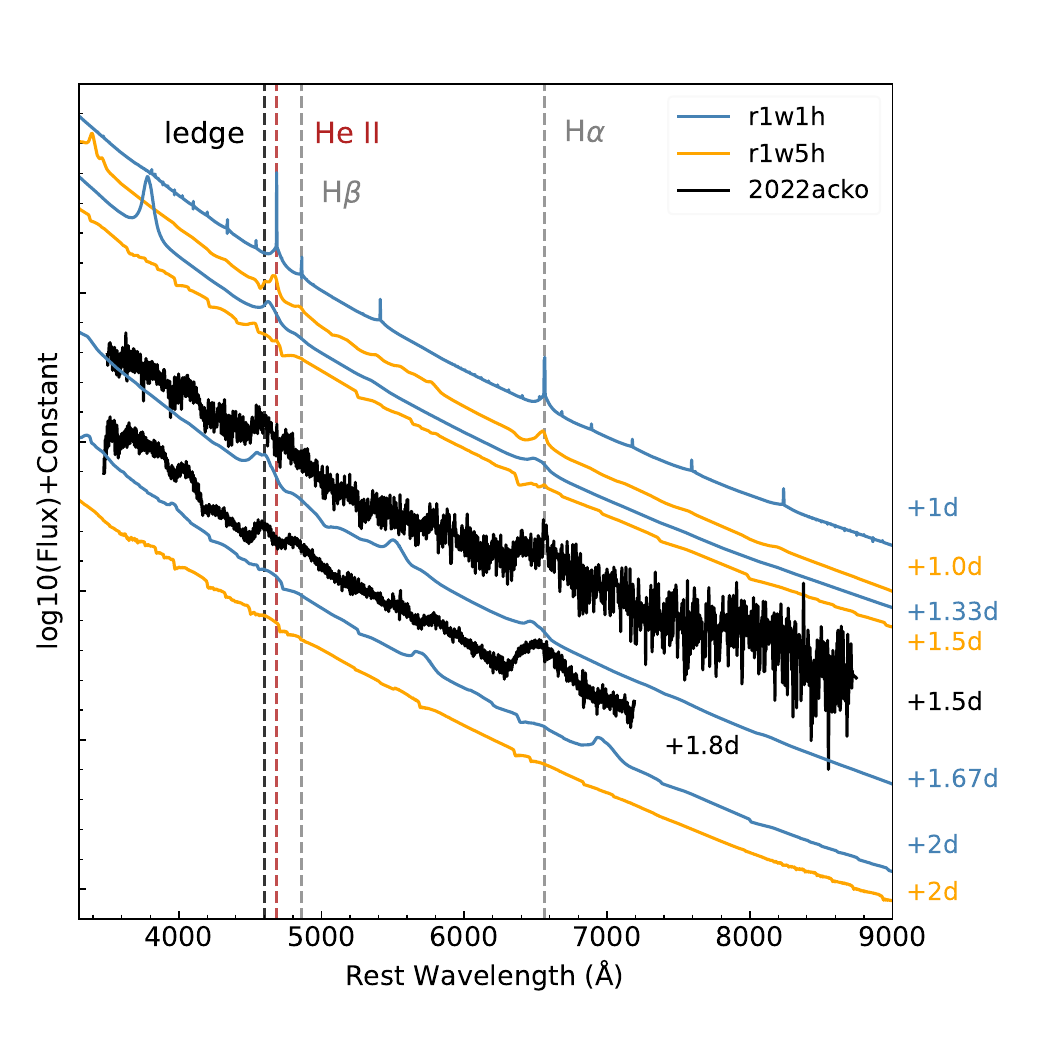}
	\caption{Comparing the early spectra of SN 2022acko to the r1w1h and r1w5h models of \citep{2017A&A...605A..83D}. Both models include an extended RSG atmosphere. The r1w1h, r1w5h model have CSM from a mass-loss rate of order $\rm 10^{-6} M_{\odot}\ yr^{-1}$ and $\rm 10^{-3} M_{\odot}\ yr^{-1}$ respectively.}
	\label{fig:r1w1h.png}
\end{figure}

\section{Discussion}

\subsection{Signs of circumstellar interaction}
\label{Signs of circumstellar interaction}

FI emission lines, produced by recombining unshocked CSM after being ionized by high energy photons created by shock-breakout or CSM-ejecta interaction, are not rare ($>30\%$ in \citealt{2023ApJ...952..119B})
 in the very early spectra of SNe II.
In normal SNe II without signs of CSI at early phases, the early spectra are characterized by the P-Cgyni profile 
of Balmer series and He~{\sc i} lines. At first glance, the early spectra of SN 2022acko are similar to that of normal SNe II.
However, we identified a broad emission (``ledge''-shaped) peaked near 4600 \AA\,at the $t = $1.5\,d spectrum of SN 2022acko. 
Moreover, we noticed a narrow emission H$\alpha$ (with an FWHM of $\sim 1100 \rm km \ s^{-1}$\footnote{
We do not say that this is the velocity of the stellar wind, since at the epoch of $t=1.5$\,d, the narrow core and the broad lorentz wings of H$\alpha$ can not be separated. 
On the other hand, as the CSM is very confined, it would be accelerated by the shock breakout radiation, leading to a much larger velocity than the stellar wind. }) superposed on the broad P-Cgyni profiles of H$\alpha$ in our $t=1.5$\,d spectrum, see Figure \ref{fig:ha_emi}. 
The ``ledge'' feature is found in dozens of SNe II, including SN II 2017eaw (see Figure \ref{fig:early_ps_compare}). There are three main interpretations for the ``ledge'' feature.  
\citet{2002AJ....124.2490L} interprets it as high velocity $\rm H\beta$. However, we do not favor this explanation since $\rm H\beta$ coexists with the ``ledge'' feature in our $t = 2.5 $\,d spectrum. 

\citet{2020ApJ...902....6S}, \citet{2023ApJ...945..107P}, \citet{2022ApJ...935...31H}, \cite{2024MNRAS.528.4209M} and \cite{2024ApJ...971..141M} suggested that ``ledge'' feature was blend of several FI emission lines, e.g., C, N and He  
in the case of SN 2018fif, SN 2018lab, SN 2021yja and SN2021gmj.
Though \citet{2022ApJ...935...31H} noticed that such interpretation (a blend of several narrow emission lines) is inconsistent with the breadth and strength of the ``ledge'' feature.
By comparing the early broad feature around 4600\AA \ to the extended atmosphere models from \cite{2017A&A...605A..83D} (i.e., r1w1h and r1w5h), although not a perfect match, 
\citet{2023ApJ...945..107P}, \citet{2022ApJ...935...31H} and \cite{2024ApJ...971..141M} proposed that an RSG with an extended atmosphere exploding into low-density CSM could be the possible scenario for such early spectra. 
In Figure \ref{fig:r1w1h.png}, we compare the early two spectra of SN 2022acko with the r1w1h and r1w5h model from \cite{2017A&A...605A..83D}, though the broad emission features near 4600\AA \ of SN 2022acko have a similar shape to that of the $t=1.0$\,d spectrum of r1w5h and the $t=1.67$\,d spectrum of r1w1h. We noticed that the shape of H$\alpha$ of SN 2022acko, a very broad P-Cgyni profile with a narrow emission superposed on it, is distinct from that of the two models. The H$\alpha$ of the $t=1.0$\,d spectrum of r1w5h and the $t=1.67$\,d spectrum of r1w1h are not that broad as seen in SN 2022acko.

\begin{figure*}
	\centering  
	\includegraphics[width=2.0\columnwidth]{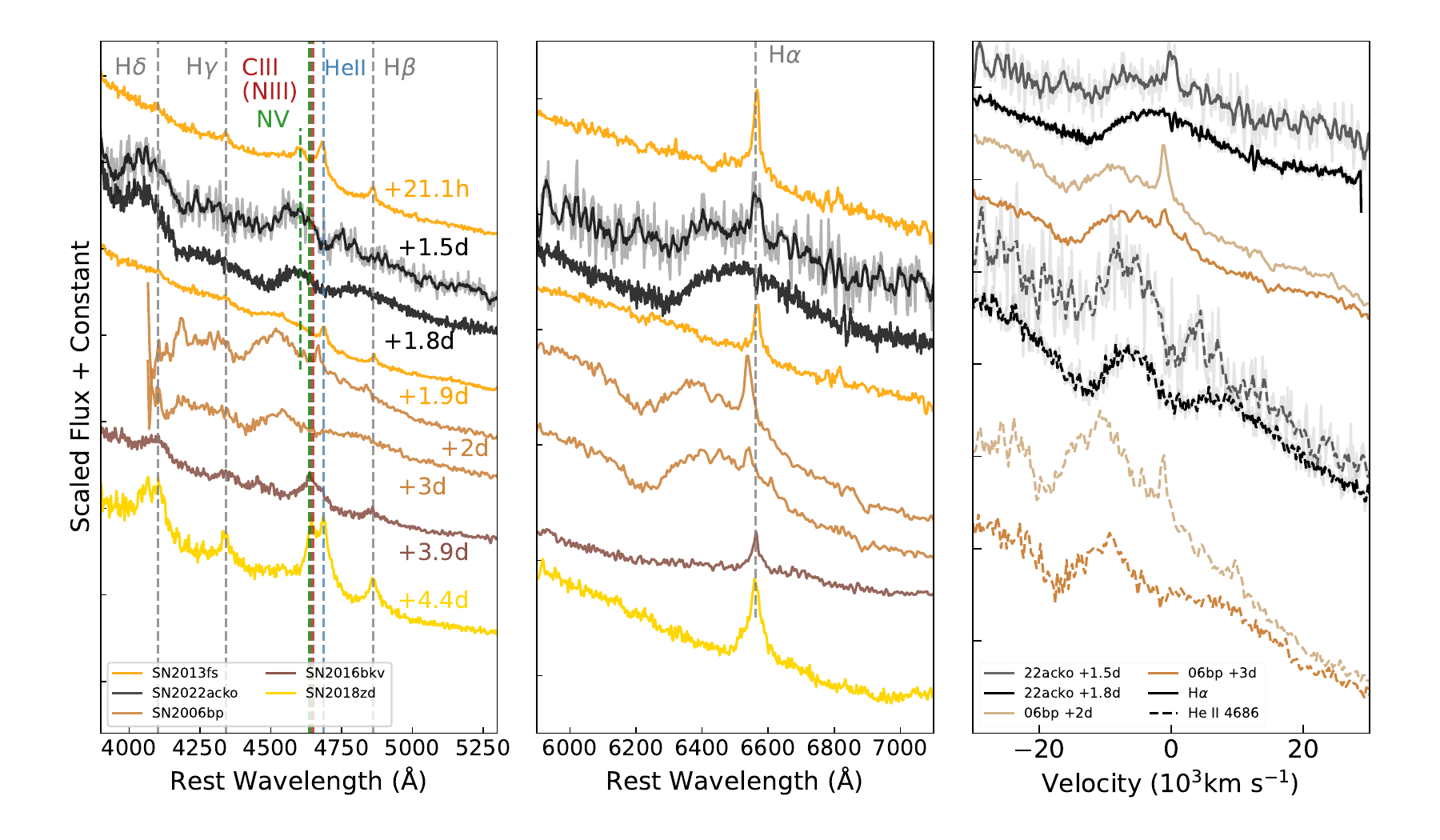}
	\caption{Left and middle panel: The region of 4000 \AA \ - 5000 \AA \ and 6000 \AA \ - 7000 \AA \ of the earlist two ($t=1.5$\,d and $t=1.8$\,d) spectra of SN 2022acko compared with Flash objects (i.e., SN 2013fs, 2018zd and SN 2016bkv) and ``ledge" object SN 2006bp. Right panel: The H$\alpha$ and He~{\sc ii} 4686 of the $t=1.5$\,d and $t=1.8$\,d spectrum of 2022acko (plotted in velocity space), compared with the $t = 2 $\,d and $t = 3$\,d spectra of SN 2006bp.} 
	\label{fig:early_compare_3}
\end{figure*}

On the other hand, in the case of SN 2006bp, \citet{2007ApJ...666.1093Q} suggested that the feature around 4600 \AA\, could be blueshifted He~{\sc ii} 4686 \AA, based on the similar shapes of He~{\sc ii} 4686 \AA\,and $\rm H\alpha$.   
\citet{2008ApJ...675..644D} also identified the ``ledge'' feature as blueshifted He~{\sc ii} in SN 2006bp, with a minor contribution of C~{\sc iii} 4650\AA\,and N~{\sc iii} 4638\AA. 
Such explanation has also been used in SN 2010id \citep{Gal-Yam_2011} and SN 2017gmr \citep{2019ApJ...885...43A}. 
In the left and middle panel of Figure \ref{fig:early_compare_3}, we compare the earliest two spectra of SN 2022acko with that of the FI objects (i.e., SN 2018zd, SN 2016bkv, and SN 2013fs) and SN 2006bp.
If the ``ledge'' feature was a blend of several flash emission lines, H$\alpha$ should also show the characteristic profile, specifically narrow emission lines with electron-scattering wings. 
However, we noticed that the H$\alpha$ of SN 2022acko was dominated by broad P-Cgyni profiles, though a weak narrow emission H$\alpha$ existed.
In the right panel of Figure \ref{fig:early_compare_3}, we compare the region of He~{\sc ii} and H$\alpha$ of the earliest two spectra of SN 2022acko with that of the SN 2006bp.
The H$\alpha$ line profile in the $t=1.5$\,d spectrum of SN 2022acko closely resembles that of SN 2006bp at $t=2$\,d. Moreover, the strong emission near He~{\sc ii} 4686 in SN 2022acko is similar to that observed in SN 2006bp 3 days after the explosion. 
In the case of SN 2006bp, the broad component around 4600 \AA \ , along with the narrow He~{\sc ii} 4686 \AA\ emission line, is reminiscent of the H$\alpha$ profile as suggested by \citet{2007ApJ...666.1093Q}. 
A comparable scenario is evident in SN 2022acko, particularly at 1.8 days after the explosion, where the broad emission around 4600\AA \ has a similar line shape to that of the H$\alpha$.
The shape of the ``ledge'' feature and the H$\alpha$ in 2022acko is more similar to that of SN 2006bp than that of the flash objects overall. 
Therefore, we suggested that the broad emission feature near 4600 \AA \ is a P-Cgyni profile of He~{\sc ii} 4686 that
comes from the {ionized SN} ejecta rather than FI blend that comes from the outermost unshocked CSM. 
The large peak blueshift indicates the outermost envelope may have a very steep density gradient that is caused by the combined actions of the shock wave passage and radiation driving at shock breakout \citep{2007ApJ...666.1093Q}.
However, we noticed that the p-cgyni profile of high ionization He~{\sc ii} 4686 is not common in the early spectra of normal SNe II (We use ``normal" to refer to SNe II that do not have CSI). 
The spectra of normal Type II Supernovae typically display He~{\sc i} 5876 lines in their early stages rather than He~{\sc ii} 4686 lines. 
When an SN explodes inside CSM, both the CSM and the SN ejecta could be ionized by high-energy photons created by CSI and produce lines of high-ionization species. The spectral lines originating from the ionized SN ejecta will exhibit distinct line shapes and velocities compared to those from the CSM.
Hence, we proposed that the broad, high-ionized, blueshifted He~{\sc ii} 4686, depicted as the "ledge" profile, is produced by the SN ejecta ionized by CSI and may therefore serve as one of the indicators of circumstellar interaction.


In the $t=1.5$\,d spectrum of SN 2022acko, we do not identify a similar narrow emission line of He~{\sc ii} 4686 {as that shown in SN 2006bp}. However, the narrow He~{\sc ii} 4686 emission line which was shown in the $t=2$\,d spectra of SN 2006bp disappeared at $t=3$\,d, see the right panel of Figure \ref{fig:early_compare_3}. We, therefore, suspected that emission lines of high-ionized species, at least for He~{\sc ii} 4686, which arise from unshocked CSM (i.e., FI features), may exist in the spectra at an epoch earlier than we observed.
The existence of narrow FI features earlier than our first spectrum could be additionally supported by \cite{2023ApJ...952..119B}. In their study of SN 2019ust, they observed an evolution from narrow He~{\sc ii} lines to a ``ledge" pattern, see their Figure 7, suggesting that the "ledge" feature emerges after the narrow FI emission lines. 
Similar transitions have been observed in SN 2013fs (see $t=1.9$\,d spectrum in Figure 2 of \citealt{2017NatPh..13..510Y}), SN 2023ixf (see the $t=5.10$\,d spectrum in extended Figure 3 of \citealt{2024Natur.627..759Z}) and SN 2024ggi (see the $t=3.42$\,d spectrum in Figure 3 of \citealt{2024ApJ...972L..15S}).
Assuming a spherical symmetric CSM, the rapid disappearing of FI features,  
which only lasts for $\sim$ 2 days, suggests that the CSM should be very confined in the vicinity of the progenitor (i.e., $\rm 2\times10^{14}\, cm$ adopting an ejecta velocity of $12000\, \rm km \ s^{-1}$).
As the narrow FI emission lines faded away, broad components from SN ejecta (e.g., the P-Cgyni of He~{\sc ii} 4686 and H$\alpha$) became dominant. 
Since the CSM layer is thin, when the ejecta rush out of the CSM, the ejecta is still in a high-ionization state and we could therefore see the ionized He~{\sc ii} features.

On the other hand, such morphology of early spectra may be related to the asymmetric structure of CSM.
\cite{2024ApJ...970..189J} suggested that departures from CSM spherical symmetry and/or homogeneous density may lead to 
Doppler-broadened line profiles and signatures of unshocked CSM appear simultaneously, which is a similar case of SN 2022acko. Assuming a disc-like or torus-like CSM environment near the stellar surface, an observer at low inclination (position C of Figure 10 in \citealt{2015MNRAS.449.1876S}) could see the symmetric narrow emission lines that arise from the CSM and the broad p-cgyni of SN ejecta at the same time. As the disc-like or torus-like CSM was quickly enveloped by the expanding SN ejecta, FI features disappeared and only a broad component of SN ejecta remains.
Future studies (e.g., obtaining high-resolution spectra capable of resolving the complete line morphology and constructing detailed spectroscopic models with various progenitor atmosphere/CSM properties) will help to uncover the origin of the ``ledge'' feature.

\subsection{Diversity of SNe II early spectra}
\begin{figure*}
	\includegraphics[width=1.5\columnwidth]{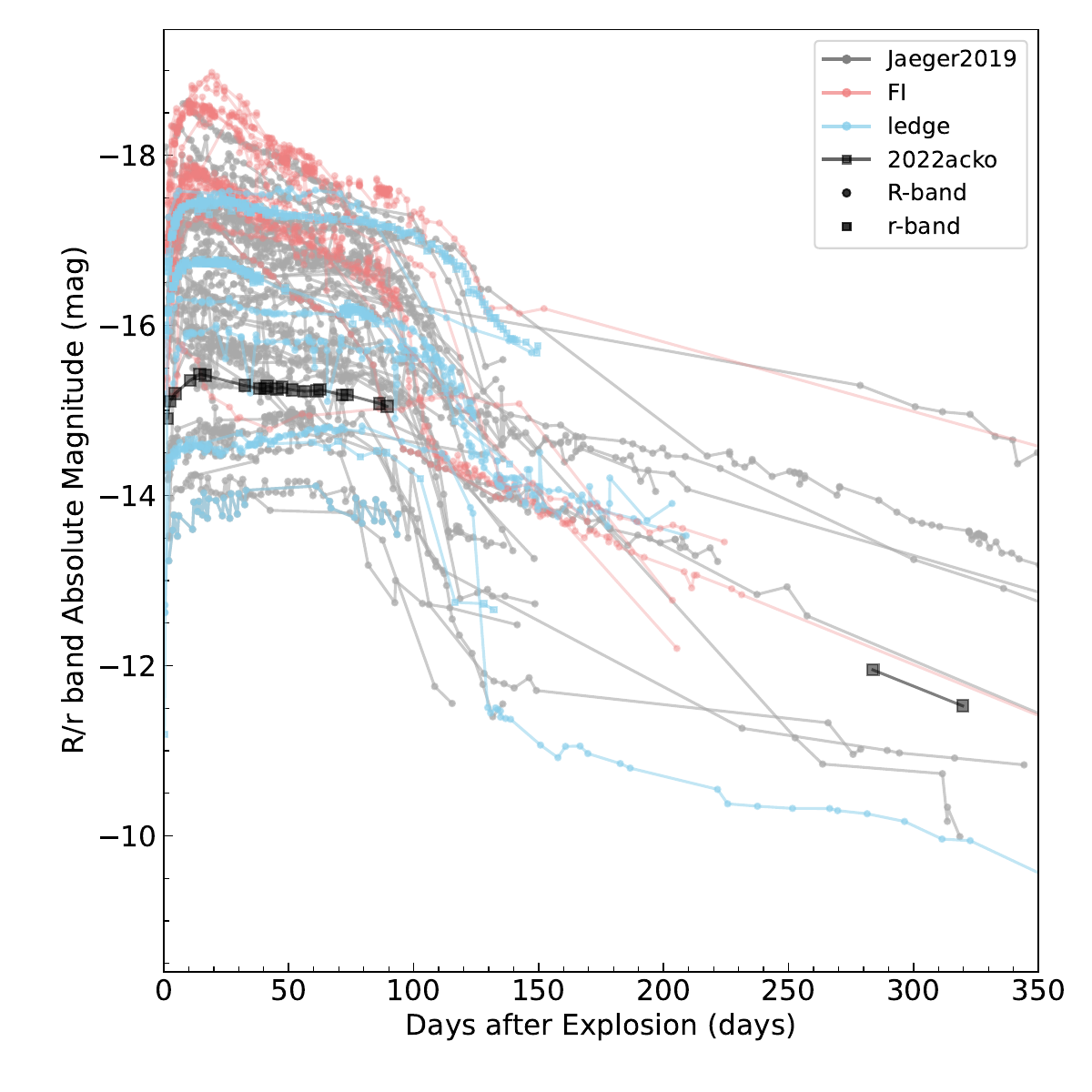}
	\caption{Comparison of the R/r-band light curve of the ``ledge'' objects (in blue) with those of Flash-ionized events (in red) and other “normal” SNe II (in gray) from \citet{2019MNRAS.490.2799D}. The FI objects were taken from \citet{2016ApJ...818....3K}, with the addition of SN 2013cu \citep{2014ApJ...789..104O}, SN 2013ej \citep{2019MNRAS.490.2799D}, SN 2014G \citep{2019MNRAS.490.2799D}, SN 2013fs \citep{2017NatPh..13..510Y}, SN 2015bf \citep{2021MNRAS.505.4890L}, SN 2016bkv, SN 2018zd. The ledge objects are SN 1999gi, SN 2005cs, SN 2010id, SN 2017eaw, SN 2017gmr, SN 2018fif, SN 2018lab, SN 2021yja, and SN 2022acko.}
	\label{fig:Rrplot}
\end{figure*}

Rapid follow-up observations have captured increasing spectra in the very early phases of SNe. These spectra are often blue and featureless (resembling a blackbody) or display FI features. In the case of SN 2022acko, the early spectra within two days post-explosion exhibited a somewhat distinct line morphology, characterized by broad P-Cygni profiles with possible narrow FI emission lines superimposed.
Noticing the spectral morphology differences between ``ledge'' objects and FI objects, we compared the R/r-band light curves of these two samples in Figure \ref{fig:Rrplot}\footnote{As discussed in section 4.1 and section 5.1, some "ledge" objects may exhibit a narrow FI emission lines before developing the distinctive broad ledge profile. Additionally, a transition from the characteristic FI emisison to ``ledge" emission has been noted in certain standard FI objects. In our study, we categorize "ledge" objects as those displaying broad emissions around 4600\AA \ but lacking or having only very weak narrow FI lines in the first spectrum, whereas FI objects are characterized by prominent FI emission lines during the early phases.}
Though \citet{2023ApJ...952..119B} found that SNe II showing FI features are not significantly brighter than those without. 
In our work, the FI samples tend to be brighter and exhibit rapidly declining light curves after reaching their peak, as suggested by \citet{2016ApJ...818....3K,2022ApJ...926...20T} and \citet{2024ApJ...970..189J}.
Conversely, the light curves of ``ledge''  samples generally decline gradually after the maximum and do not typically appear in the high-luminosity range. 
Given that there are no or only very weak and extremely rapid disappearance of narrow emission lines in ``ledge'' objects, we suggested that CSM in ``ledge'' objects should be confined in the very vicinity of the progenitor and the circumstellar interaction has a moderate or less contribution to the whole light curve than that of the FI sample. 

\cite{2024ApJ...970..189J} classified the FI samples into three groups based on their early-time spectra: Class 1 shows high-ionization lines of He~{\sc ii}, N~{\sc iii}, and C~{\sc iv}, Class 2 exhibits emission lines of He~{\sc ii} and C~{\sc iv} but lacks N~{\sc iii}, and Class 3 shows only weak, narrow He~{\sc ii} narrow emission overlaying on a blueshifted, Doppler-broadened He~{\sc ii}. They attributed this diversity to variations in the extent and density of CSM. However, it should be noted that such classification is epoch-dependent. 
In \citet{2024ApJ...970..189J}'s work, most of the ledge objects (e.g., SN 2017eaw, SN 2017gmr, SN 2018lab, SN 2021gmj, SN 2021yja) have been classified as comparison objects with no CSI, suggesting that the CSI in ``ledge" objects are weak and can be easily ignored.
If narrow He~{\sc ii} 4686 was detected, then SN 2022acko and other ``ledge" objects would be classified into group 3, the same category as SN 2013fs. 
However, there is evidence that ledge objects are distinct from Class 3 of \citet{2024ApJ...970..189J}:
the H$\alpha$ of ``ledge" objects are dominated by broad p-cgyni profiles whereas H$\alpha$ in Class 3 objects (including SN 2023ixf, 2024ggi, and SN 2019ust) still exhibit FI emission lines with electron-scattering wings.

By deriving mass-loss rates and CSM densities for 39 SNe II with FI features and 35 SNe II without such spectral signatures, \cite{2024ApJ...970..189J} proposed a continuum of mass-loss histories in the final years to months before the explosion, assuming all these objects have an RSG progenitor. 
The non-existing or extremely rapid-disappearance of narrow FI features in the spectra and the very confined CSM near the progenitor surface make ``ledge" objects appear to
be intermediate between FI objects and those without signs of CSM interaction. 
Conversely, the ``ledge" phenomenon could be linked to a particular mechanism, such as aspherical CSM, which results in a distinct spectral morphology different from typical FI spectra.

%

\subsection{Explosion parameters}

\subsubsection{Nickel Mass}
The nickel mass of SN 2022acko can be estimated from its late-time luminosity, as the luminosity during the nebular phase is powered by the $\rm ^{56}Ni$ and its decay product $\rm ^{56}Co$. For SN 2022acko, it is reasonable to assume all the gamma rays due to $\rm ^{56}Co$ to $\rm ^{56}Fe$ are entirely trapped, given that the decay rate of the late-time luminosity matches the radioactive decay rate. The nickel mass of SN 2022acko was estimated to be $\rm 0.015\pm0.005M_{\odot}$ from the second equation of \cite{2003ApJ...582..905H}. 
The nickel mass can also be derived by comparing the luminosity of SN 2022acko to that of SN 1987A. The mass of $\rm ^{56}Ni$ of SN 1987A has been determined to be 0.075 $\pm$ 0.005$\rm M_{\odot}$ \citep{1996snih.book.....A}. The $L(\rm 2022acko)$/$L(\rm 1987A)$ is $0.246\pm0.078$ at 284 days after the explosion, from which we estimated the nickel mass of SN 2022acko to be $0.018^{+0.008}_{-0.006}$, consist with that estimated from \citep{2003ApJ...582..905H}. 
Therefore, we adopt the average value of $\rm 0.017^{+0.009}_{-0.007}M_{\odot}$ as the mass of nickel for SN 2022acko.

\subsubsection{Comparing with Hydrodynamical Lightcurve Models}

In order to relate the observational characteristics of SN 2022acko to its explosion properties,
we compared the bolometric light curve, temperature evolution, and velocity evolution of SN 2022acko to a grid of hydrodynamical models constructed by \citet{2022ApJ...934...67B} and \citet{2023ApJ...952..155Z}. 
Both studies initiate with non-rotating, solar-metallicity stellar progenitors from \cite{2016ApJ...821...38S}.
\citet{2022ApJ...934...67B} simulated the core collapse and subsequent explosion in 1D with the FLASH code \citep{2000ApJS..131..273F} incorporating the STIR model \citep{2020ApJ...890..127C}, whereas \cite{2023ApJ...952..155Z} employ a semi-analytical approach based on \citet{2016MNRAS.460..742M}.
Despite employing quite different implementations and
degrees of approximations, \cite{2023ApJ...952..155Z} found considerable agreements with the results (explosion energy and baryonic neutron star mass) achieved by \citet{2022ApJ...934...67B} except for the mass ranges with near-critical explodability. 
Radioactive heating from the $\rm ^{56}Ni \to ^{56}Co \to ^{56}Fe$ decay chain and the mixing extent of nickel 
play crucial roles in powering and influencing the light-curve properties.
None of these two studies adopted the accurate $\rm ^{56}Ni$ mass which can only be obtained by multi-D neutrino-transport simulations.
Instead, \citet{2023ApJ...952..155Z} employ their Equation 1 for a rough estimation of nickel mass combining shock-heated and neutrino-heated production, and then spread $\rm M_{Ni}$ homogeneously up to $\rm 3 M_{\odot}$. On the other hand,  \citet{2022ApJ...934...67B} utilize the relation of $\rm ^{56}Ni$ yields and explosion energy from \citet{2016ApJ...821...38S} to estimate the
$\rm M_{Ni}$ and disperse the $\rm ^{56}Ni$ up to about 75\% of the way through the He shell. 
Subsequently, to obtain the light curves, \citet{2023ApJ...952..155Z} employed the thermal bomb method with the supernova explosion code (SNEC; \citealt{2015ApJ...814...63M}) and the explosion properties (e.g., explosion energy and nickel mass), while \citet{2022ApJ...934...67B} mapped their simulated explosion profiles directly into SNEC.

As shown from the top panel of Figure \ref{fig:lcmodel}, for the same progenitor, the theoretical bolometric luminosity, velocity and temperature predicted by  \citet{2023ApJ...952..155Z} exceed those of \citet{2022ApJ...934...67B} across all phases. This reflects that the explosion energy and nickel mass of \citet{2023ApJ...952..155Z} are a factor of $\sim2$ larger than those of \citet{2022ApJ...934...67B} in the considered mass range here, i.e., $9-11\,M_{\sun}$. 
At the nebular phase, the theoretical bolometric luminosity of \citet{2022ApJ...934...67B} underestimated the flux while the models of \citet{2023ApJ...952..155Z} predict a good agreement with that of SN 2022acko, suggesting that at least for SN 2022acko, the relatively simple approach for nickel production used in \citet{2023ApJ...952..155Z} is effective.

In our comparison with theoretical models, we focused mainly on the properties during the plateau phase i.e., from $\sim 30$ to $\sim 100$ days after the explosion, as circumstellar interaction and nickel mixing exhibit minimal influence during this period. 
While none of the theoretical bolometric light curves align closely with that of SN 2022acko, the bolometric luminosity of SN 2022acko at the plateau phase is closest to those produced by a low-mass progenitor, specifically in the range of 9-10\,$\rm M_{\odot}$.)
The photospheric velocity of SN 2022acko between 60 and 100 days after the explosion is similar to that of 9-10\,$\rm M_{\odot}$ progenitors, whereas between 30 to 60 days post explosion, the photospheric velocity is more matched with that of 10-12\,$\rm M_{\odot}$ progenitors.
The evolution of blackbody temperature in SN 2022acko at the plateau phase is closest to that of 9-10\,$\rm M_{\odot}$ progenitors as well. However, as the data point errors encompass the full spread of models, it is challenging to differentiate the progenitor mass based on temperature evolution alone.
Additionally, \citet{2019ApJ...879....3G} suggested that velocities measured during the majority of the plateau phase are highly correlated with bolometric luminosity, indicating limited additional insights into the explosion properties. 
We thus mainly rely on the comparison of bolometric lightcurve to estimate the explosion parameters of the SN 2022acko.
The explosion energy, ejecta mass, and progenitor radius for these 9-10\,$\rm M_{\odot}$ models range from 2.6-4.5$\rm \times 10^{50} erg$, $\rm \sim 7 M_{\odot}$, and 400-500\,$\rm R_{\odot}$ respectively.

\subsubsection{Hybrid Analytic Model for Light-Curve}

\begin{figure*}
	
	\subfigure{
		\includegraphics[width=1.8\columnwidth]{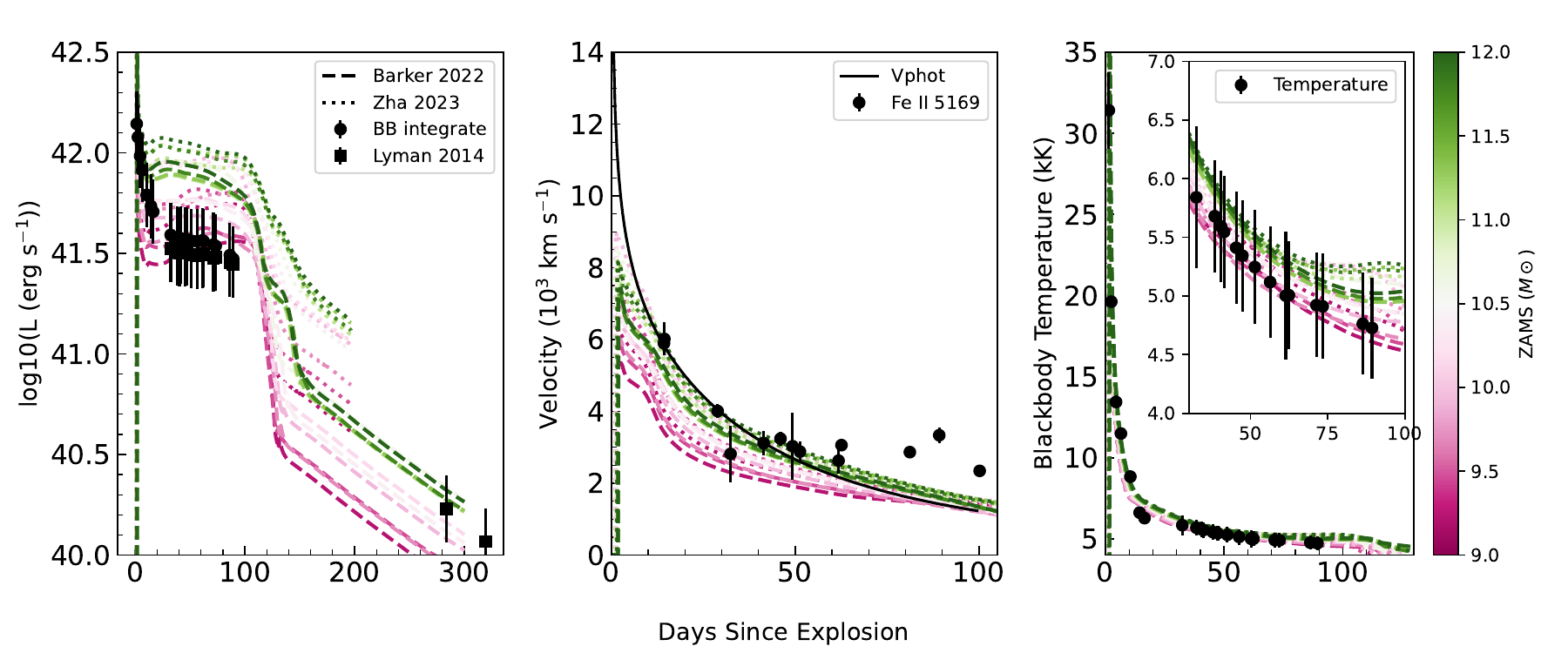}
	}
	
	\subfigure{
		\includegraphics[width=1.8\columnwidth]{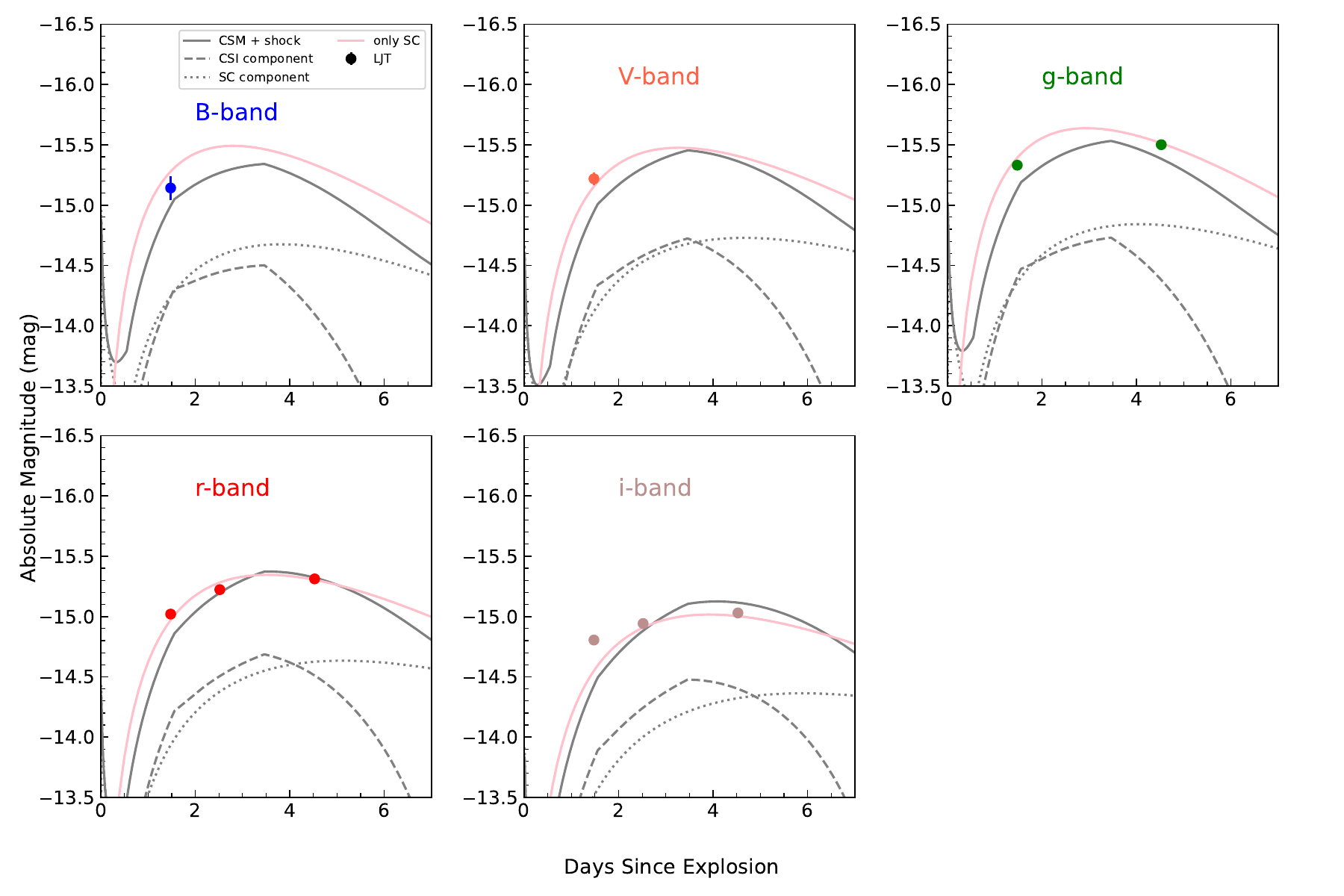}
	}	
	\caption{Top three panels: Comparison between the observed properties (i.e., bolometric light curves, temperature evolution, and velocity evolution) of SN 2022acko with the theoretical expectation produced by an RSG with an initial mass between $9\sim 12 \rm M_{\odot}$. The theoretical models are taken from \citet{2022ApJ...934...67B} (dashed line) and \citet{2023ApJ...952..155Z} (dotted line), respectively. Lower five panels: Fitting the early multiband light curve of SN 2022acko with a multiple-process hybrid model. The best-fit hybrid model (shock cooling plus CSM interaction) is shown in a gray solid line.} 
	\label{fig:lcmodel}
\end{figure*}

The explosion parameters were also estimated by other methods.
\citet{2016ApJ...829..109M} found that the properties of the light curve at early phases depend mainly on the progenitor radius and they thus provide a relation between the g-band rise time and the radius at the time of explosion. Based on this relation (Equation 4 of \citet{2016ApJ...829..109M}), the progenitor radius at the time of the explosion is $\sim 533 \rm R_{\odot}$.
Before the nebular phases, the radiation of SNe II is dominated by shock breakout and subsequent cooling. 
Various studies have explored analytical models to understand the shock cooling emission. 
In this work, we first utilized the shock breakout theory developed by \cite{2010ApJ...725..904N} and ran a Markov Chain Monte Carlo (MCMC) code to fit the early light curve. 
However, when considering only the shock cooling emission, the estimated values for explosion energy, progenitor mass and progenitor radius were $\rm 5^{+3.2}_{-2.5}\times10^{51}\,erg$, $\rm 15^{+5}_{-5}M_{\odot}$ and $\rm 250^{+120}_{-60}R_{\odot}$, see the pink line in the lower panel of Figure \ref{fig:lcmodel}. 
These values appear inconsistent with those expected for a LLSN. 
Given that the early spectra of SN 2022acko show signs of interaction
(i.e., the broad "ledge" feature and narrow H$\alpha$ emission lines),
we utilized the hybrid model proposed by \citet{2024Natur.627..754L}, which incorporates the radiation from shock cooling and CSM interaction, to match the early-time multiband (BVgri) light curve of SN 2022acko. 
The shock cooling emission was modeled following \citet{2010ApJ...725..904N},
while the CSI process is described based on the frameworks established by \citet{2013MNRAS.435.1520M} and \citet{2024arXiv241106351H}. To describe the confined CSM surrounding SN 2022acko, we adopt the CSM density profiles used in \citet{2024Natur.627..754L}.
(for a detailed description of this model, see \cite{2024Natur.627..754L}). 
As the circumstellar interaction of SN 2022acko lasts for merely a few days, we restrict the photometry data analysis within 6 days after the explosion for fitting purposes.
Guided by our prior knowledge of the approximate explosion parameter (i.e., explosion energy, ejecta mass and progenitor radius) range and our expertise in light curve fitting, we manually adjust the parameters and see how good the fit is. 
The best matched of the multiband optical light curves of SN 2022acko suggests a confined CSM ($\sim 2\times10^{14}$\,cm), resulting from a mass loss of $\rm \sim 5\times 10^{-4} \rm M_{\odot} \ year^{-1}$ (assuming csm expansion velocity is $\rm 50\,km\ s^{-1}$) shortly before the SN explosion. The explosion energy, ejecta mass, and radius of the progenitor are estimated to be $\rm \sim 0.5\times 10^{51} erg$, $\rm \sim 5 M_{\odot}$, $\rm \sim 400 R_{\odot}$, respectively. Such combination of explosion parameters is consistent with the expectations from LLSNe.


\section{Conclusions}
In this work, we have presented high-cadence, early photometric, and spectroscopic observations, as well as late-time photometry of SN II 2022acko. 
With an peak absolute V-band magnitude of $M_V= -15.5 \pm 0.3$ mag and mid-plateau expansion velocities of $v_{\rm FeII 5169}^{50}=2756\pm 210 \rm \ km\ s^{-1}$, SN 2022acko falls into the category of low-luminosity SNe IIP. 
Its photometric properties closely resemble those of SN 2008in and SN 2009N. The $uvw2$ magnitude of SN 2022acko peaks at $\sim 18$\,mag at 1.7\,d after the explosion.  In contrast to some SNe II that show an r-band rise during the plateau phase, the r-band light curve of SN 2022acko shows a monotonic decline after its peak. The bolometric luminosity of SN 2022acko reaches a peak luminosity of $\rm 1.4\times 10^{42} erg\ s^{-1}$ immediately after the explosion and enters into the plateau with a bolometric luminosity of $\rm 4.8\times 10^{41} erg\ s^{-1}$ 25 days after the explosion. We additionally estimated the UV flux ratio for 2022acko, the UV flux accounts for 73\% of the total luminosity at +1.48\,d after the explosion while this fraction decreases to 5\% at +16.5\,d after the explosion.

The spectral evolution of SN 2022acko is overall typical among the SNe II sample. 
However, it shows some peculiarities. The broad "ledge" feature at 4600\AA \ and the narrow H$\alpha$ with a FWHM of $\rm \sim1100\,km\ s^{-1}$ makes SN 2022acko among the rare LL sample with early CSI. 
Moreover, the Ba~{\sc ii} \ 6497 line that often shows in LLSNe is not detected (at least not detected in 100 days after the explosion) and the Fe line forest was not detected either. 
We suggested that the non-detection of Ba~{\sc ii} 6497 and Fe forest may be attributed to the higher expansion velocities during the second half of the plateau, causing Ba~{\sc ii} 6497 and H$\alpha$, as well as Fe line forest and H$\beta$, to merge. 
In this work, we calculated the metallicity (from the N2 indicator) for 17 LLSNe, and we found that the distribution of metallicities of the LLSNe environment is uniform spread. The average metallicity of the LLSN environment is $8.6\pm0.2$. 
Compared to other low-luminosity SNe II, SN 2022acko 
has a relative high metallicity.

By directly comparing the observed properties (light curve, temperature evolution, and velocity evolution) of SN 2022acko with that of a synthetic set of SNe IIP models, we suggested that SN 2022acko should have a low mass ($\rm M_{ZAMS} \sim$9-10\,$\rm M_{\odot}$) progenitor, 
This speculation is consistent with the suggestion of a low-mass progenitor (i.e., less than 8 $\rm M_{\odot}$) derived from pre-explosion archival Hubble Space Telescope images in \cite{2023MNRAS.524.2186V}.
The nickel mass is found to be  $\rm 0.017^{+0.009}_{-0.007}M_{\odot}$ from the tail luminosity. 
Based on our hybrid light curve model (shocking cooling + CSM interaction) and comparison with synthetic set of SNe IIP models, 
the explosion energy, the ejecta mass and radius of the progenitor were estimated as $\sim$0.26-0.5$\rm \times 10^{51} erg$, $\sim$ 5-7\,$\rm M_{\odot}$ and $\sim $ 400-500\,$\rm R_{\odot}$ respectively.
From our hybrid light curve model, we also estimated a mass loss rate of $\sim 5 \times 10^{-4} \rm M_{\odot} \ year^{-1}$ shortly before the SN explosion and a CSM outer radius of $\rm \sim 2\times 10^{14}\, cm $.

At first glance, the early spectra of SN 2022acko show a broad P-Cgyni profile of the H$\alpha$, which is a common feature among SNe II. 
However, we observed a distinct broad ``ledge'' emission near 4600 \AA\,in the earliest two spectra of SN 2022acko. We interpret this feature as blueshifted He~{\sc ii} emission originating from the SN ejecta ionized by CSI.
The large blueshift of the peak may result from the steep density gradient in the outermost envelope caused by the shock wave and shock breakout passage. 
Additionally, a narrow H$\alpha$ line (with an FWHM of $\rm \sim 1100\,km\ s^{-1}$), 
was detected superimposed on a broad H$\alpha$ component in the 1.5\,d spectrum but quickly disappeared in the 1.8\,d spectrum. We proposed that this narrow H$\alpha$ line is the FI emission line generated by the recombination of the unshocked CSM.  
Based on the similar early spectral features with SN 2006bp and a transition from narrow He~{\sc ii} FI lines to a broad structure in some strong FI objects (SN 2019ust, SN 2013fs, SN 2013ixf and SN 2024ggi), we hypothesized that the narrow emission line of He~{\sc ii} may also occur at an earlier stage than we detected. 
Assuming spherical CSM, the rapid disappearance of the FI emission lines (H$\alpha$ and possible He~{\sc ii}) suggests that CSM is confined at the very vicinity of progenitor (i.e., $\sim \rm 2\times10^{14}\, cm$). 
Despite being estimated using different methods, the CSM radius obtained from the narrow FI feature matches that derived from the light curve modeling.
As the SN ejecta move through the thin CSM layer, they 
remain highly ionized, leading to the observable broad, ionized emission of He~{\sc ii} 4686. 
The broad ``ledge" structure has also been observed in several other SNe II, including the typical SNe II 2017eaw. In these ledge objects, the narrow FI emission lines ahead of the ledge feature were easy to miss.
Compared with the standard FI samples, the ``ledge" sample appears to have a more confined CSM environment, 
potentially representing an intermediate class bridging the FI objects and normal SNe II. 
The more constrained CSM environment in "ledge" objects is further supported by the absence of redward evolution in the uvw2-v color. 
Moreover, as suggested by \cite{2024ApJ...970..189J}, the unique spectral morphology observed in SN 2022acko, i.e., the simultaneous detection of the p-cgyni profile of fast-moving SN ejecta and FI emission lines of outermost unshocked CSM, could be linked to an aspherical CSM.  
We caution that other explanations, e.g., the interaction between the extended envelope of an RSG progenitor with low-density CSM, blends of fading FI lines, have also been proposed to explain the ``ledge'' feature. Future synthetic spectra models incorporating different properties of progenitor atmosphere and/or CSM
could help uncover the origin of the ``ledge'' features.



 

\section*{Acknowledgements}
We would like to thank the anonymous referee for constructive comments which have improved the work immensely.
This work is supported by the National Key R\&D Program of China with No. 2021YFA1600404.
H.Lin is supported by the National Natural Science Foundation of China (NSFC grants No. 12403061)
and the innovative project of ``Caiyun Postdoctoral Project" of Yunnan Province.
This work is supported by the National Natural Science Foundation of China (12173082, 12333008), the Yunnan Fundamental Research Projects (grants 202501AV070012, 202401BC070007 and 202201AT070069), the Topnotch Young Talents Program of Yunnan Province, the Light of West China Program provided by the Chinese Academy of Sciences, the International Centre of Supernovae, Yunnan Key Laboratory (No. 202302AN360001) and the science research grants from the China Manned Space Project with No. CMS-CSST-2021-A12. 
X.Wang is supported by the National Natural Science Foundation of China (NSFC grants 12288102, 12033003, and 11633002), the Scholar Program of Beijing Academy of Science and Technology (DZ:BS202002), and the Tencent Xplorer Prize.
M. Hu is supported by the Postdoctoral Fellowship Program of CPSF under Grant Number GZB20240376 and the Shuimu Tsinghua Scholar Program. S. Zha is supported by the National Natural Science Foundation of China (Grant Nos. 12473031, 12393811) and the Yunnan Fundamental Research Projects (Grant No. 202501AS070078).
Y.-Z. Cai is supported by the National Natural Science Foundation of China (NSFC, Grant No. 12303054) and the Yunnan Fundamental Research Projects (Grant No. 202401AU070063).
X.Z. acknowledges financial support from the Natural Science Foundation of the Jiangsu Higher Education Institutions of China (No.24KJB160001) and the scientific research fund of Jiangsu Second Normal University (928201/059).
AR acknowledges financial support from the GRAWITA Large Program Grant (PI P. D’Avanzo) and from the PRIN-INAF 2022 "Shedding light on the nature of gap transients: from the observations to the models".
W.L. Lin is supported by the National Natural Science Foundation of China (NSFC grants12473045, 12033003, 12494572), and the Natural Science Foundation of Xiamen, China (grant 3502Z202471015).
We acknowledge the support of the staff of the LJT. Funding for the LJT has been provided by the CAS and the People’s Government of Yunnan Province. The LJT is jointly operated and administrated by YNAO and the Center for Astronomical Mega-Science, CAS.

\section*{Data Availability}

The data underlying this article are available in the article and its online supplementary material.



\bibliographystyle{mnras}
\bibliography{mnras_template} 




\appendix

\section{Basic information and observational data of SN 2022acko}
\begin{figure*}
	\includegraphics[width=0.9\columnwidth]{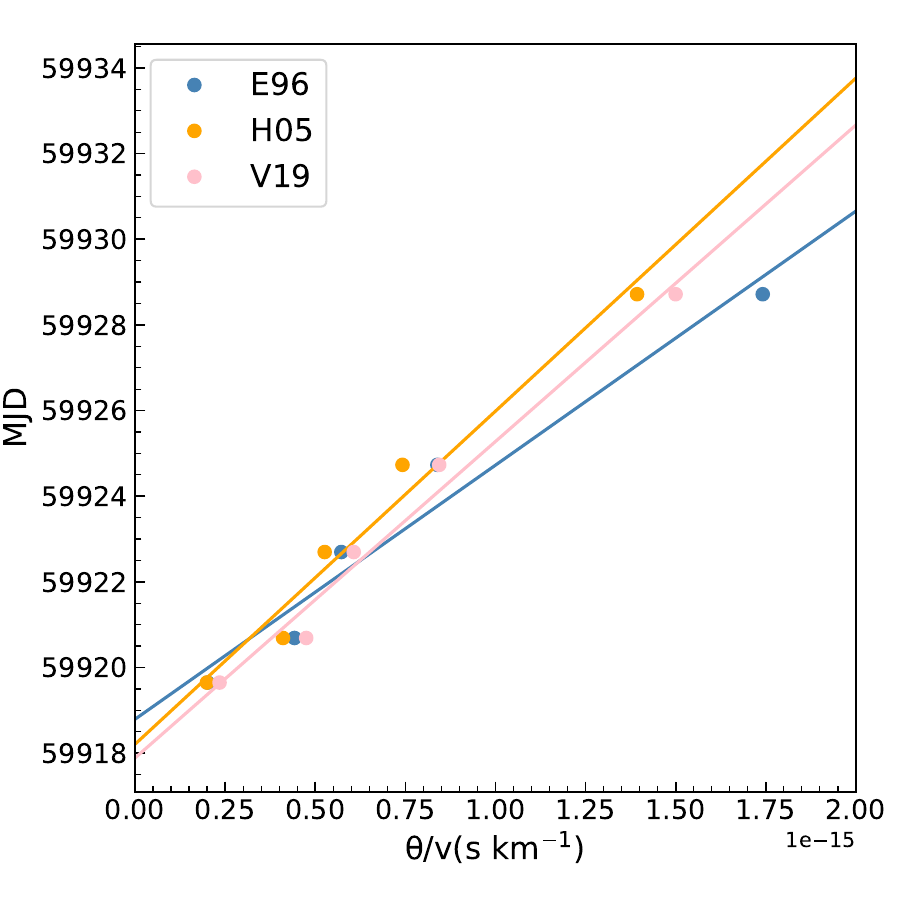}
	\caption{MJD as a function of $\theta /v$ for different form of dillusion factor, with slope for distance and ordinate crossing at explosion time. }
	\label{fig:epm}
\end{figure*}

\begin{figure*}
	\includegraphics[width=1.5\columnwidth]{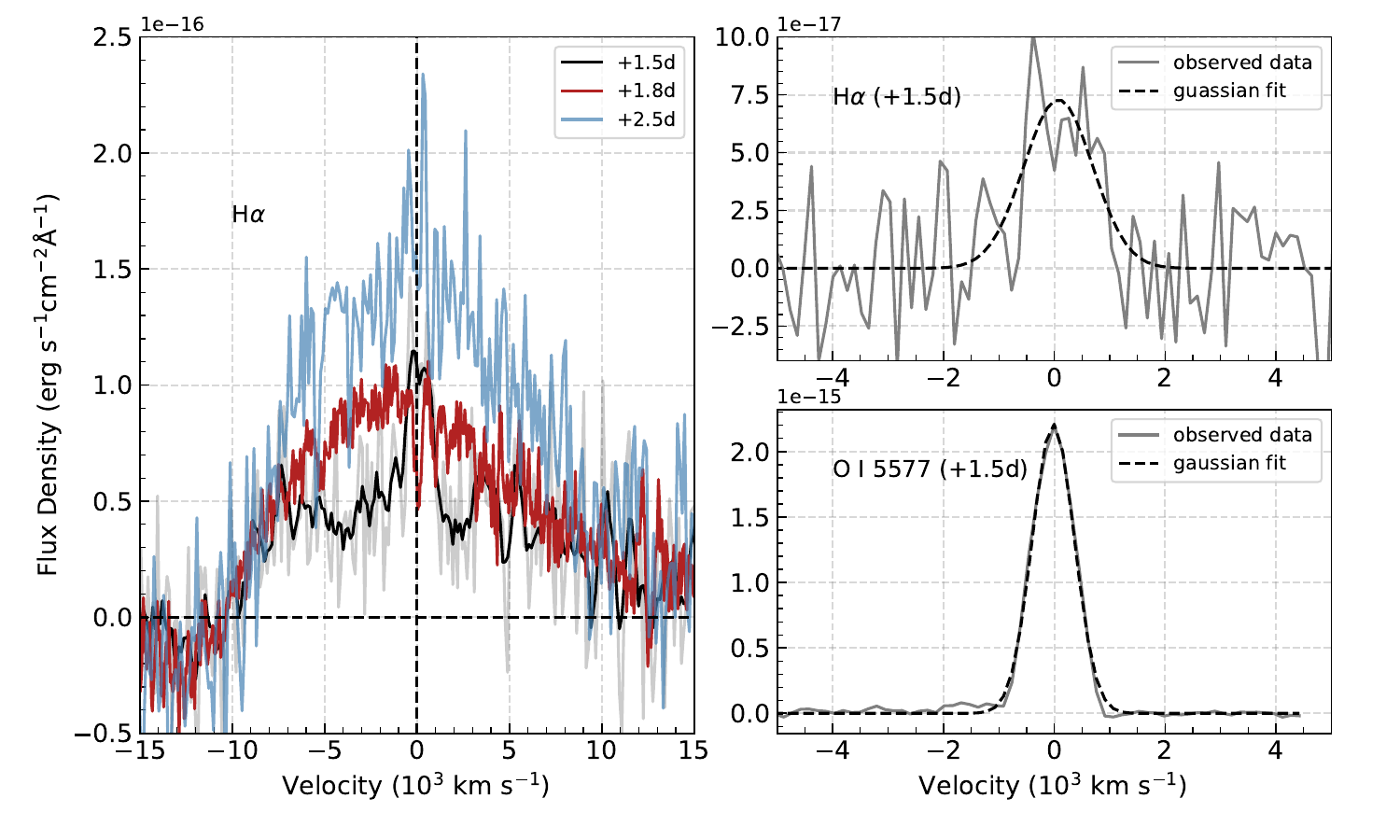}
	\caption{Left Panel: the evolution of H$\alpha$ of SN 2022acko at early phases. A narrow emission line of H$\alpha$ (see the black dashed line), which can be clearly seen at the $t = $ 1.5\,d spectrum, however, can not be detected at the $t = $ 1.8\,d spectrum and is not evident at the $t = $ 2.5\,d. Right panel: H$\alpha$ (upper) and O~{\sc i} 5577 (lower) sky emission lines for the $t = $ 1.5\,d spectrum of SN 2022acko. The FWHM of this narrow H$\alpha$ emission line is $\rm \sim 1390 \pm 400 \, km\ s^{-1}$. The FWHM of O~{\sc i} 5577 line, which can be treated as the instrumental FWHM, is $\rm 840\, km\ s^{-1}$. After the instrumental correction was applied by $\rm FWHM_{cor} = (FWHM_{obs}^{2}-FWHM_{inst}^{2})^{1/2}$, the FWHM of narrow H$\alpha$ is $\rm \sim 1100 \pm 550 \, km\ s^{-1}$.}
	\label{fig:ha_emi}
\end{figure*}

In this section, we provide the basic information (Table \ref{tab:B_I} and Table \ref{tab:distance}) and the observational data of SN 2022acko (Table \ref{tab:opt_phot},Table \ref{tab:swift} and Table \ref{tab:spectra}) and its standard stars (Table \ref{tab:standardstar}). Besides, we provide the metallicity of LLSNe sample (Table \ref{tab:m_llsne}), the additional information about the EPM method (Table \ref{tab:para_epm} and Figure \ref{fig:epm}) and the H$\alpha$ evolution at early phases (Figure \ref{fig:ha_emi}).

\begin{table}
	\centering
	\caption{Basic information of SN 2022acko}
	\label{tab:B_I}
	\begin{tabular}{ll}
		\hline
		Property & Value \\
		\hline
		RA                        & 03:19:38.990                         \\
		DEC                       & -19:23:42.68                         \\
		Host                      & NGC 1300                   \\
		redshift                  & 0.00526                    \\
		explosion date            & 59918.17                   \\
		$E(B-V)_{\rm MW}$                  & 0.026                      \\
		$E(B-V)_{\rm host}$               & 0.03 $\pm$ 0.01$\rm ^a$        \\
		\hline
	\end{tabular}
	\begin{tablenotes}
		\item $\rm ^aE(B-V)$ is taken from \cite{2023ApJ...953L..18B}, however we adopt $Rv=1.4$ for host galaxy.		
	\end{tablenotes}
\end{table}	

\begin{table*}
	\centering
	\caption{Quantities that derived and used in EPM.}
	\label{tab:para_epm}
        \begin{tabular}{llllllllll}
        \hline
        MJD & Phase (d) & T (K)$\rm ^a$ & $v_{\rm phot} (\rm km\ s^{-1})$ &$\xi_{\rm E96}$ & $\theta_{\rm E96}\ (\rm 10^{-12}rad)$ & $\xi_{\rm H05}$ & $\theta_{\rm H05}\ (\rm 10^{-12}rad)$ & $\xi_{\rm V19}$ & $\theta_{\rm V19}\ (\rm 10^{-12}rad)$ \\
        \hline
        59919.65 &+1.48& 31421$\pm$2367 & 11315 & 0.534 & 2.28 & 0.539 & 2.26 & 0.459 & 2.65 \\
        59920.69 &+2.52& 19620$\pm$256  & 9969  & 0.474 & 4.41  & 0.510 & 4.09 & 0.441 & 4.74 \\
        59922.70 &+4.53& 18613$\pm$2053 & 8564  & 0.467 & 4.90  & 0.508 & 4.51 & 0.441 & 5.20 \\
        59924.73 &+6.56& 15470$\pm$1424 & 7692  & 0.445 & 6.45  & 0.503 & 5.71 & 0.442 & 6.49 \\
        59928.72 &+10.55& 10131$\pm$594  & 6568  & 0.424 & 11.43 & 0.530 & 9.15 & 0.493 & 9.85 \\
        \hline
        \end{tabular}
        \begin{tablenotes}
		\item $\rm ^a$ In MJD 59919.65 and 59920.69, the temperature was derived using photometry of $uvw2$, $uvw1$, $u$, B, V, $g$, $r$ and $i$ bands. In MJD 59922.70, 59924.73 and 59928.72, the temperature was derived using data of optical band.
	\end{tablenotes}
\end{table*}	

\begin{table}
	\centering
	\caption{Distance of SN 2022acko with various method}
	\label{tab:distance}
	\begin{tabular}{lll}
		\hline
	    Distance &  Method  &  Reference   \\
		\hline
		14.993$\pm$4.29\,Mpc  &  Truly-fisher & NED   \\
		18.99$\pm$2.85\,Mpc   &  Numerical action method  & \cite{2021MNRAS.501.3621A} \\
		16.61$\pm$3.32\,Mpc    &  EPM &\citet{1996ApJ...466..911E}   \\
		21.78$\pm$4.34\,Mpc    &  EPM &\citet{2005AA...439..671D}   \\
		20.69$\pm$4.14\,Mpc    &  EPM &\citet{2019AA...621A..29V}   \\
		21.97$\pm$1.98\,Mpc    &  SCM &\citet{2002ApJ...566L..63H}   \\
		23.4$\pm$3.9\,Mpc      &  SCM &\citet{2023MNRAS.524.2186V}   \\	
		19.8$\pm$2.84\,Mpc	   &  average &                         \\
        20.69$\pm$4.14\,Mpc & median & \\ 
		\hline
	\end{tabular}
\end{table}		

\begin{table*}
	\centering
	\caption{metallicity of LLSNe sample}
	\label{tab:m_llsne}
	\begin{threeparttable}
	\begin{tabular}{lllllll}
		\hline
		SN       & Host Galaxy    & $Z_{\rm host \ center}$ & $Z_{\rm SN \ loc}$ & $Z_{\rm sn \ loc} [Z\odot]$ & Comments  &       spectrum information               \\
        \hline  
		SN1999br   & NGC4900        & $\cdots$            & 8.55$^1$   & 0.724     &      &                           \\
		SN2002gd   & NGC7537        & $\cdots$            & 8.58$^1$   & 0.776     &      &                           \\
		SN2003Z    & NGC2742        & $\cdots$            & 8.66$^2$   & 0.933     &      &                           \\
		SN2005cs   & M51            & $\cdots$            & 8.66$^3$   & 0.933     &      &                           \\
		SN2006ov   & M61            & $\cdots$            & 8.9$^4$    & 1.621     &      &                           \\
		SN2008bk   & NGC7793        & $\cdots$            & 8.58$^1$   & 0.776     &      &                           \\
		SN2008in   & NGC4303        & $\cdots$            & 8.88$^1$   & 1.548     &      &                           \\
		SN2009N    & NGC4487        & $\cdots$            & 8.57$^1$   & 0.758     &      &                          \\
		SN2009md   & NGC3389        & $\cdots$            & 8.96$^5$   & 1.862     &      &                           \\
		SN2016aqf  & NGC2101        & $\cdots$            & 8.134$^6$  & 0.278     &      &                           \\
		SN2016bkv  & NGC3184        & 8.645$\rm ^a$           & 8.58$\rm ^b$   & 0.776     & $N_{2,\rm hc}$:-0.447,$R_{\rm sn}/R_{25}$=0.1312 &  \citet{2010MNRAS.405..735R}$\rm ^c$   \\
		SN2018hwm  & Ic2327         & 8.5$^7$             & 8.30$\rm ^b$   & 0.407     & $R_{\rm sn}/R_{25}$ = 0.42   &               \\
		SN2018lab  & NGC2207/Ic2163 & $\cdots$            & 8.64$\rm ^a$   & 0.891     & $N_{2,\rm sn}$:-0.451     &    2018-12-30; SALT+RSS; \citet{2023ApJ...945..107P}          \\
		SN2020cxd  & NGC6359        & $\cdots$            & 8.5$^8$    & 0.646     &      &                           \\
		SN2021aai  & NGC2268        & 8.791$\rm ^a$           & 8.72$\rm ^b$   & 1.071     & $N_{2,\rm hc}$:-0.191, $R_{\rm sn}/R_{25}$ = 0.1498   &  \citet{1995ApJS...98..477H}$\rm ^c$   \\
		SN2021gmj  & NGC3310        & $\cdots$            & 8.55$\rm ^a$   & 0.724     & $N_{2,\rm sn}$=-0.61    &  2021-03-21; KOOLS-IFU; \cite{2024MNRAS.528.4209M}               \\
		SN2022acko & NGC1300        & 8.886$\rm ^a$           & 8.81$\rm ^a$   & 1.318     & $N_{2,\rm hc}$:-0.024   $N_{2,\rm sn}$:-0.157       &  2022-12-10; Bok+B\&C; \cite{2023ApJ...953L..18B}         \\
		\hline  
	\end{tabular}
    \begin{tablenotes}
    	\item $\rm ^a$Metallicity calculated from $\rm N_{2}$ indicator.
    	\item $\rm ^b$Metallicity calculated using the average metallicity gradient.
            \item $\rm ^c$The host spectrum is found from the NED. 
    	\item  1-8 Metallicity taken from literature, with
        [1]\citet{2016A&A...589A.110A}, [2]\citet{2022MNRAS.512.1541G}, [3]\citet{2005MNRAS.364L..33M}, [4]\citet{2011MNRAS.410.2767C}, [5]\citet{2011MNRAS.417.1417F}, [6]\citet{2020MNRAS.497..361M},
        [7]\citet{2021MNRAS.501.1059R}, [8]\citet{2021A&A...655A..90Y}
    \end{tablenotes}
    \end{threeparttable}
\end{table*}

\begin{table*}
	\centering
	\caption{Reference Stars in the SN 2022acko Field}
	\label{tab:standardstar}
	\begin{tabular}{llllllll} 
		\hline
		Num & $\alpha$(J2000) & $\delta$(J2000) & $B$ (mag) & $V$ (mag) & $g$ (mag) & $r$ (mag) & $i$ (mag) \\
		\hline
		1   & 49.874319 & -19.357571 & 17.379(0.099) & 15.879(0.008) & 16.439(0.006)  & 15.333(0.003)  & 14.327(0.002) \\
		2   & 49.874116 & -19.387435 & 16.665(0.081) & 15.748(0.06)  & 16.058(0.003)  & 15.537(0.004)  & 15.344(0.003) \\
		3   & 49.89214  & -19.39345  & 16.781(0.149) & 15.715(0.033) & 16.129(0.0003) & 15.503(0.0007) & 15.238(0.003) \\
		4   & 49.967612 & -19.39952  & 16.571(0.082) & 15.838(0.064) & 16.074(0.006)  & 15.647(0.002)  & 15.511(0.004) \\
		5   & 49.993439 & -19.431067 & 16.357(0.036) & 15.303(0.029) & 15.701(0.001)  & 14.98(0.002)   & 14.69(0.003)  \\       
		\hline
	\end{tabular}
\end{table*}

\begin{table*}
	\centering
	\caption{Optical photometry of SN 2022acko}
	\label{tab:opt_phot}	
	\begin{tabular}{llllllll}
		\hline
		MJD        & phase (d)  & B            & V           & g            & r            & i             & Telescope \\
		\hline
		59919.6481 & +1.48   & 16.548(0.099) & 16.405(0.053) & 16.325(0.025) & 16.576(0.027) & 16.756(0.036) & LJT       \\
		59920.6887 & +2.52   & $\cdots$      & $\cdots$      & $\cdots$      & 16.373(0.016) & 16.619(0.019) & LJT       \\
		59922.6959 & +4.53   & $\cdots$      & $\cdots$      & 16.155(0.01)  & 16.284(0.007) & 16.531(0.008) & LJT       \\
		59924.7315 & +6.56   & $\cdots$      & $\cdots$      & 16.061(0.01)  & 16.17(0.008)  & $\cdots$      & LJT       \\
		59928.7163 & +10.55  & 16.431(0.097) & 16.112(0.044) & 16.149(0.007) & 16.128(0.006) & 16.347(0.008) & LJT       \\
		59932.6012 & +14.43  & 16.527(0.097) & 16.064(0.044) & 16.155(0.007) & 16.055(0.007) & 16.256(0.007) & LJT       \\
		59934.6744 & +16.5   & 16.65(0.097)  & 16.094(0.044) & 16.237(0.006) & 16.07(0.004)  & 16.249(0.005) & LJT       \\
		59950.5793 & +32.41  & 17.355(0.099) & 16.336(0.045) & 16.684(0.014) & 16.189(0.009) & 16.236(0.017) & LJT       \\
		59956.5414 & +38.37  & $\cdots$      & $\cdots$      & 16.775(0.007) & 16.226(0.006) & 16.227(0.006) & LJT       \\
		59958.6160 & +40.45  & $\cdots$      & $\cdots$      & 16.806(0.009) & 16.225(0.006) & 16.226(0.007) & LJT       \\
		59959.5583 & +41.39  & $\cdots$      & $\cdots$      & 16.793(0.009) & 16.196(0.006) & 16.207(0.006) & LJT       \\
		59963.4944 & +45.32  & $\cdots$      & $\cdots$      & 16.851(0.01)  & 16.228(0.006) & 16.212(0.006) & LJT       \\
		59965.5202 & +47.35  & $\cdots$      & $\cdots$      & 16.875(0.008) & 16.208(0.006) & 16.209(0.006) & LJT       \\
		59969.5382 & +51.37  & $\cdots$      & $\cdots$      & 16.937(0.007) & 16.243(0.006) & 16.228(0.006) & LJT       \\
		59974.5061 & +56.34  & $\cdots$      & $\cdots$      & 16.997(0.01)  & 16.257(0.006) & 16.226(0.007) & LJT       \\
		59979.5490 & +61.38  & 17.878(0.101) & 16.506(0.045) & 17.014(0.014) & 16.25(0.008)  & 16.226(0.01)  & LJT       \\
		59980.5722 & +62.4   & $\cdots$      & $\cdots$      & 17.022(0.015) & 16.237(0.008) & 16.211(0.009) & LJT       \\
		59989.5073 & +71.34  & $\cdots$      & $\cdots$      & 17.115(0.01)  & 16.3(0.006)   & 16.25(0.007)  & LJT       \\
		59991.5459 & +73.38  & $\cdots$      & $\cdots$      & 17.144(0.011) & 16.298(0.007) & 16.269(0.006) & LJT       \\
		60004.5084 & +86.34  & $\cdots$      & $\cdots$      & 17.352(0.019) & 16.401(0.011) & 16.38(0.01)   & LJT       \\
		60007.5117 & +89.34  & $\cdots$      & $\cdots$      & $\cdots$      & 16.435(0.012) & 16.426(0.014) & LJT       \\
		60201.9052 & +283.74 & $\cdots$      & $\cdots$      & 20.627(0.041) & 19.532(0.021) & 19.732(0.033) & LJT       \\
		60237.8056 & +319.64 & $\cdots$      & $\cdots$      & 21.081(0.074) & 19.955(0.032) & 20.019(0.046) & LJT    \\
		\hline   
	\end{tabular}
\end{table*}

\onecolumn

\begin{longtable}{lllll}
	\caption{Swift UVOT photometry of SN 2022acko}   \\
	\label{tab:swift}  \\ 
	\hline
	MJD        & Phase (d) $^a$ & Magnitude & $\sigma$ & Filter    \\
	\hline	
	\endfirsthead	
	\multicolumn{5}{c}{Swift UVOT photometry of SN 2022acko (Continue)} \\
	\hline	
	MJD        & Phase (d) $^a$ & Magnitude & $\sigma$ & Filter  \\
	\hline	
	\endhead	
	
	\hline
	\endfoot
	
	\hline
	\multicolumn{5}{l}{$^a$Phase relative to the estimated explosion date.} \\	
	\endlastfoot
	59919.8727 & +1.70      & 14.302   & 0.028    & uvw2   \\
	59920.3385 & +2.17      & 14.512   & 0.027    & uvw2   \\
	59920.7394 & +2.57      & 14.668   & 0.027    & uvw2   \\
	59921.3863 & +3.22      & 14.897   & 0.030    & uvw2   \\
	59922.4519 & +4.28      & 15.272   & 0.032    & uvw2   \\
	59922.7828 & +4.61      & 15.386   & 0.034    & uvw2   \\
	59923.1055 & +4.94      & 15.420   & 0.033    & uvw2   \\
	59923.4494 & +5.28      & 15.563   & 0.034    & uvw2   \\
	59924.3678 & +6.20      & 15.807   & 0.039    & uvw2   \\
	59924.6322 & +6.46      & 15.836   & 0.039    & uvw2   \\
	59925.4242 & +7.25      & 16.080   & 0.038    & uvw2   \\
	59926.7588 & +8.59      & 16.511   & 0.049    & uvw2   \\
	59927.4194 & +9.25      & 16.480   & 0.059    & uvw2   \\
	59928.1445 & +9.97      & 16.871   & 0.063    & uvw2   \\
	59929.7486 & +11.58     & 17.344   & 0.082    & uvw2   \\
	59930.4645 & +12.29     & 17.524   & 0.096    & uvw2   \\
	59931.4638 & +13.30     & 18.003   & 0.107    & uvw2   \\
	59932.5882 & +14.42     & 18.420   & 0.158    & uvw2   \\
	59933.3234 & +15.15     & 18.213   & 0.132    & uvw2   \\
	59934.6442 & +16.47     & 18.267   & 0.145    & uvw2   \\
	59919.6662 & +1.5       & 14.320   & 0.039    & uvm2   \\
	59919.8774 & +1.71      & 14.354   & 0.028    & uvm2   \\
	59920.3409 & +2.17      & 14.401   & 0.028    & uvm2   \\
	59920.7429 & +2.57      & 14.437   & 0.029    & uvm2   \\
	59921.3923 & +3.22      & 14.620   & 0.029    & uvm2   \\
	59922.4580 & +4.29      & 14.837   & 0.030    & uvm2   \\
	59922.7889 & +4.62      & 14.905   & 0.030    & uvm2   \\
	59923.1087 & +4.94      & 15.090   & 0.031    & uvm2   \\
	59923.4529 & +5.28      & 15.146   & 0.032    & uvm2   \\
	59924.3739 & +6.2       & 15.438   & 0.035    & uvm2   \\
	59924.6384 & +6.47      & 15.537   & 0.036    & uvm2   \\
	59925.4264 & +7.26      & 15.885   & 0.037    & uvm2   \\
	59926.7631 & +8.59      & 16.358   & 0.045    & uvm2   \\
	59927.4237 & +9.25      & 16.606   & 0.066    & uvm2   \\
	59928.1475 & +9.98      & 16.901   & 0.065    & uvm2   \\
	59929.7542 & +11.58     & 17.172   & 0.229    & uvm2   \\
	59930.4700 & +12.3      & 17.656   & 0.105    & uvm2   \\
	59931.4669 & +13.3      & 18.196   & 0.119    & uvm2   \\
	59932.5928 & +14.42     & 18.218   & 0.130    & uvm2   \\
	59933.3262 & +15.16     & 18.260   & 0.130    & uvm2   \\
	59934.6482 & +16.48     & 18.473   & 0.169    & uvm2   \\
	59919.8688 & +1.7       & 14.544   & 0.034    & uvw1   \\
	59920.3365 & +2.17      & 14.629   & 0.030    & uvw1   \\
	59920.7364 & +2.57      & 14.586   & 0.030    & uvw1   \\
	59921.3812 & +3.21      & 14.630   & 0.031    & uvw1   \\
	59922.4467 & +4.28      & 14.779   & 0.032    & uvw1   \\
	59922.7776 & +4.61      & 14.775   & 0.032    & uvw1   \\
	59923.1027 & +4.93      & 14.884   & 0.032    & uvw1   \\
	59923.4465 & +5.28      & 14.889   & 0.031    & uvw1   \\
	59924.3626 & +6.2       & 15.137   & 0.035    & uvw1   \\
	59924.6270 & +6.46      & 15.203   & 0.036    & uvw1   \\
	59925.4223 & +7.25      & 15.429   & 0.035    & uvw1   \\
	59926.7551 & +8.59      & 15.721   & 0.042    & uvw1   \\
	59927.4157 & +9.25      & 15.957   & 0.054    & uvw1   \\
	59928.1419 & +9.97      & 16.179   & 0.053    & uvw1   \\
	59929.7437 & +11.58     & 16.473   & 0.063    & uvw1   \\
	59930.4599 & +12.29     & 16.866   & 0.080    & uvw1   \\
	59931.4611 & +13.29     & 17.005   & 0.074    & uvw1   \\
	59932.5844 & +14.42     & 17.569   & 0.115    & uvw1   \\
	59933.3210 & +15.15     & 17.466   & 0.099    & uvw1   \\
	59934.6408 & +16.47     & 17.751   & 0.127    & uvw1   \\
	59920.3376 & +2.17      & 15.004   & 0.038    & u      \\
	59920.7381 & +2.57      & 14.980   & 0.038    & u      \\
	59921.3842 & +3.21      & 14.946   & 0.038    & u      \\
	59922.4497 & +4.28      & 14.983   & 0.039    & u      \\
	59922.7807 & +4.61      & 15.004   & 0.039    & u      \\
	59923.1043 & +4.93      & 14.990   & 0.038    & u      \\
	59923.4482 & +5.28      & 14.955   & 0.036    & u      \\
	59924.3657 & +6.2       & 15.027   & 0.039    & u      \\
	59924.6301 & +6.46      & 15.040   & 0.039    & u      \\
	59925.4234 & +7.25      & 15.100   & 0.036    & u      \\
	59926.7573 & +8.59      & 15.244   & 0.041    & u      \\
	59927.4178 & +9.25      & 15.217   & 0.047    & u      \\
	59928.1434 & +9.97      & 15.325   & 0.044    & u      \\
	59929.7466 & +11.58     & 15.484   & 0.047    & u      \\
	59930.4626 & +12.29     & 15.700   & 0.053    & u      \\
	59931.4626 & +13.29     & 15.881   & 0.049    & u      \\
	59932.5866 & +14.42     & 16.129   & 0.060    & u      \\
	59933.3224 & +15.15     & 16.277   & 0.062    & u      \\
	59934.6428 & +16.47     & 16.577   & 0.077    & u      \\
	59919.3289 & +1.16      & 16.460   & 0.055    & b      \\
	59919.6605 & +1.49      & 16.323   & 0.051    & b      \\
	59919.7923 & +1.62      & 16.429   & 0.054    & b      \\
	59919.8161 & +1.65      & 16.316   & 0.051    & b      \\
	59919.8717 & +1.7       & 16.269   & 0.049    & b      \\
	59920.3381 & +2.17      & 16.285   & 0.046    & b      \\
	59920.7388 & +2.57      & 16.311   & 0.046    & b      \\
	59921.3853 & +3.22      & 16.340   & 0.049    & b      \\
	59922.4508 & +4.28      & 16.212   & 0.046    & b      \\
	59922.7817 & +4.61      & 16.258   & 0.047    & b      \\
	59923.1049 & +4.93      & 16.167   & 0.044    & b      \\
	59923.4488 & +5.28      & 16.189   & 0.043    & b      \\
	59924.3667 & +6.2       & 16.271   & 0.047    & b      \\
	59924.6311 & +6.46      & 16.255   & 0.047    & b      \\
	59925.4238 & +7.25      & 16.249   & 0.041    & b      \\
	59926.7580 & +8.59      & 16.282   & 0.045    & b      \\
	59927.4186 & +9.25      & 16.313   & 0.056    & b      \\
	59928.1439 & +9.97      & 16.238   & 0.047    & b      \\
	59929.7476 & +11.58     & 16.299   & 0.050    & b      \\
	59930.4635 & +12.29     & 16.401   & 0.053    & b      \\
	59931.4632 & +13.29     & 16.338   & 0.044    & b      \\
	59932.5874 & +14.42     & 16.475   & 0.052    & b      \\
	59933.3229 & +15.15     & 16.476   & 0.050    & b      \\
	59934.6435 & +16.47     & 16.546   & 0.054    & b      \\
	59919.3337 & +1.16      & 16.545   & 0.105    & v      \\
	59919.6652 & +1.5       & 16.611   & 0.111    & v      \\
	59919.7970 & +1.63      & 16.511   & 0.104    & v      \\
	59919.8764 & +1.71      & 16.481   & 0.101    & v      \\
	59920.3405 & +2.17      & 16.415   & 0.091    & v      \\
	59920.7422 & +2.57      & 16.374   & 0.089    & v      \\
	59921.3913 & +3.22      & 16.313   & 0.089    & v      \\
	59922.4570 & +4.29      & 16.268   & 0.087    & v      \\
	59922.7878 & +4.62      & 16.216   & 0.083    & v      \\
	59923.1081 & +4.94      & 16.212   & 0.080    & v      \\
	59923.4523 & +5.28      & 16.206   & 0.078    & v      \\
	59924.3729 & +6.2       & 16.147   & 0.081    & v      \\
	59924.6374 & +6.47      & 16.360   & 0.091    & v      \\
	59925.4260 & +7.26      & 16.280   & 0.075    & v      \\
	59926.7624 & +8.59      & 16.165   & 0.076    & v      \\
	59927.4230 & +9.25      & 16.187   & 0.098    & v      \\
	59928.1469 & +9.98      & 16.169   & 0.083    & v      \\
	59929.7533 & +11.58     & 16.223   & 0.088    & v      \\
	59930.4690 & +12.3      & 16.257   & 0.092    & v      \\
	59931.4663 & +13.3      & 16.123   & 0.070    & v      \\
	59932.5920 & +14.42     & 16.269   & 0.086    & v      \\
	59933.3257 & +15.16     & 16.159   & 0.077    & v      \\
	59934.6475 & +16.48     & 16.086   & 0.076    & v      \\
	
\end{longtable}

\twocolumn

\begin{table*}
	\centering
	\caption{Log of Spectroscopic Observations of SN 2022acko}
	\label{tab:spectra}
	\begin{threeparttable}
	\begin{tabular}{lllcccc} 
		\hline
		UT Date & JD & Phase (d) $^a$  & Telescope & Instrument & Range (\AA) \\
		\hline
        20221206 & 59919.67 & +1.5  & LJT & YFOSC+G3+slit1.8 & 3500-8800   \\
        20221207 & 59920.66 & +2.5  & LJT & YFOSC+G3+slit1.8 & 3500-8800   \\
        20221211 & 59924.71 & +6.5  & LJT & YFOSC+G3+slit2.5 & 3500-8800   \\
        20221219 & 59932.57 & +14.4 & LJT & YFOSC+G3+slit2.5 & 3500-8800   \\
        20230106 & 59950.55 & +32.4 & LJT & YFOSC+G3+slit2.5 & 3600-8900   \\
        20230115 & 59959.53 & +41.4 & LJT & YFOSC+G3+slit2.5 & 3600-8900   \\
        20230123 & 59967.52 & +49.3 & LJT & YFOSC+G3+slit2.5 & 3600-8900   \\
        20230125 & 59969.50  & +51.3 & LJT & YFOSC+G3+slit2.5 & 3600-8900   \\
        20230205 & 59980.53  & +61.8 & LJT & YFOSC+G3+slit2.5 & 3600-8900   \\
		\hline
	\end{tabular}
    \begin{tablenotes}
    	\item $^a$Phase relative to the estimated explosion date.
    \end{tablenotes}
    \end{threeparttable}
\end{table*}

\bsp	
\label{lastpage}
\end{document}